Dear Reader,

below you find the LaTeX source of our paper TUM-T31-60/94 on K_L -->
Pi^0 e+e-. We have tried to make the processing of the LaTeX source and
PostScript figures as convenient as possible:

 * Get the set of 9 PostScript figures belonging to this paper from the
   bulletin board and unpack them.
 * Just put everything below the `--- snip ---' in a separate LaTeX file
   in the same directory where you have unpacked the PostScript figures
   before. Run your LaTeX processor on this file.
 * LaTeX will then ask you interactively whether you have the required
   style files `epsf.sty' and `rotate.sty' (available separately as macros
   at this bulletin board) to include the accompanying PostScript figures
   with your `dvips' converter (of course having these style files doesn't
   help if your version of "dvips" can not handle them).
 * If you have any problems concerning the inclusion of PostScript figures
   simply answer the first question with `No'. This then gives you at least
   a version of the paper without figures.

There should be no need for you to edit the LaTeX file as far as the
treatment of PostScript figures is concerned.

Best regards,
MARKUS LAUTENBACHER

---
MARKUS E. LAUTENBACHER,         | phone: +49/89/3209-2387
Technical University Munich,    | fax:   +49/89/32092296
Physics Dept.-Theo. Physics T31,| Internet:
D-85748 Garching, Germany       | lauten@feynman.t30.physik.tu-muenchen.de

--- snip --- snip --- snip --- snip --- snip --- snip --- snip --- snip ---
\typein[\YourChoice]{}

\def\empty{}
\ifx \YourChoice \empty
   \renewcommand{\YourChoice}{0}
\fi

\ifcase \YourChoice
   \documentstyle[12pt]{article}
\or
   \documentstyle[12pt,epsf,rotate]{article}
\else
   \renewcommand{\YourChoice}{0}
   \documentstyle[12pt]{article}
\fi

\ifnum \YourChoice=1
   \typein[\YourChoice]{}

   \def\empty{}
   \ifx \YourChoice \empty
      \renewcommand{\YourChoice}{1}
   \else
      \ifcase \YourChoice
         \renewcommand{\YourChoice}{1}
         \or \or
      \else
         \renewcommand{\YourChoice}{1}
      \fi
   \fi
\fi

\newcommand{\figintext}[1]{
\ifnum \YourChoice=1
   #1
\fi
}

\newcommand{\figatend}[3]{
\ifcase \YourChoice
   #1
   #3
\or
\or
   #1
   #2
   #3
\fi
}

\newcommand{\newsection}[1]{\section{#1}\setcounter{equation}{0}}

\newcommand{\be}{\begin{equation}}
\newcommand{\ee}{\end{equation}}
\newcommand{\bd}{\begin{displaymath}}
\newcommand{\ed}{\end{displaymath}}
\newcommand{\f}{\frac}
\newcommand{\ra}{\rightarrow}
\newcommand{\me}[1]{\langle#1\rangle}
\newcommand{\al}{\alpha_s}
\newcommand{\aem}{\alpha}
\newcommand{\kpiee}{$K_L \ra \pi^0 e^+ e^-$ }
\newcommand{\Lms}{\Lambda_{\overline{\rm MS}}}

\newcommand{\mt}{m_{\rm t}}

\newcommand{\gev}{\, {\rm GeV}}
\newcommand{\mev}{\, {\rm MeV}}

\newcommand{\IM}{{\rm Im}}

\begin{document}


\renewcommand{\thefootnote}{\fnsymbol{footnote}}

\author{
{Andrzej J. BURAS${}^{1,2}$, \qquad Markus E. LAUTENBACHER${}^{1}$}\\
{Miko\l aj MISIAK${}^{1}$\thanks{On leave of absence from Institute of
Theoretical Physics, Warsaw University.
},
\qquad\phantom{Xx} Manfred M\"UNZ${}^{1}$
\phantom{XXXXXXXXl}}\\
{\small\sl ${}^{1}$ Physik Department, Technische Universit\"at
M\"unchen, D-85748 Garching, Germany.}\\
{\small\sl ${}^{2}$ Max-Planck-Institut f\"ur Physik
                    -- Werner-Heisenberg-Institut,}\\
{\small\sl F\"ohringer Ring 6, D-80805 M\"unchen, Germany.}
}

\date{}

\title{
{\large\sf
\rightline{MPI-Ph/94-11}
\rightline{TUM-T31-60/94}
\rightline{February 1994}
}
\bigskip
\bigskip
{\LARGE\sf
Direct CP Violation in \kpiee Beyond Leading
Logarithms}\footnote{Supported by the German Bundesministerium f\"ur
Forschung und Technologie under contract 06 TM 732, the CEC
Science project SC1-CT91-0729 and the Polish Committee for Scientific
Research.
}
}

\maketitle
\thispagestyle{empty}

\begin{abstract}
\noindent
We analyze the direct CP violation in the rare decay \kpiee with QCD
effects taken into account consistently in the next-to-leading order.
We calculate the two-loop mixing between the four-quark $\Delta S=1$
operators and the operator $Q_{7V} =
(\overline{s}d)_{V-A}(\overline{e}e)_V$ in the NDR and HV
renormalization schemes. Using the known two-loop anomalous dimension
matrix of the four-quark operators, we find that the coefficient
$C_{7V}(\mu)$ depends only very weakly on $\mu$, renormalization
scheme and $\Lms$. The next-to-leading QCD corrections enhance the
direct CP violating contribution over its leading order estimate so
that it remains dominant in spite of the recent decrease of
$|V_{ub}/V_{cb}|$ and $|V_{cb}|$. We expect typically $BR(K_L \ra
\pi^0 e^+ e^-)_{dir} \approx 6\cdot 10^{-12}$, although values
as high as $10^{-11}$ are not yet excluded.
\end{abstract}

\newpage
\setcounter{page}{1}

\setcounter{footnote}{0}
\renewcommand{\thefootnote}{\arabic{footnote}}

\newsection{Introduction} \label{s.introd}

        A clear cut observation of {\em direct} CP violation, the
violation of CP symmetry in the decay amplitudes remains as one of the
central targets of high energy physics in the 90's and in the
beginning of 21st century
\cite{winsteinwolfenstein:93,ritchiewojcicki:93,littenbergvalencia:93}.
The CP violation in the $K\ra\pi\pi$ decays discovered almost 30 years
ago can be accommodated in the standard model as a result of complex
phases present in $K^0$--$\overline{K}^0$ mixing and is usually called
the {\em indirect} CP violation.

        The Kobayashi-Maskawa description of CP violation
\cite{kobayashimaskawa:73} predicts also the existence of this
phenomenon in the decay amplitudes with its size strongly correlated
with the weak couplings of the top quark and with its mass. In this
respect, a very special role is played by the ratio
$\epsilon'/\epsilon$. A measurement of a non-zero
$Re(\epsilon'/\epsilon)$ would automatically signal a {\em direct} CP
violation. The experimental situation is not conclusive however. While
the result of NA31 collaboration at CERN with $Re(\epsilon'/\epsilon)
= (23 \pm 7)\cdot 10^{-4}$ \cite{wagner:93} clearly indicates
direct CP violation, the value of E731 at Fermilab,
$Re(\epsilon'/\epsilon) = (7.4 \pm 5.9)\cdot 10^{-4}$
\cite{gibbons:93} is compatible with superweak theories in which
$\epsilon'/\epsilon = 0$. Hopefully, in about five years the
experimental situation concerning $\epsilon'/\epsilon$ will be
clarified through the improved measurements by the two collaborations
at the $10^{-4}$ level and by experiments at the $\Phi$ factory in
Frascati.

        On the theoretical side a considerable progress has been made
by calculating the short distance Wilson coefficients of the operators
contributing to $\epsilon'/\epsilon$ beyond leading logarithmic
approximation
\cite{altarellietal:81,burasweisz:90,burasetal:92a,burasetal:92b,burasetal:92c,burasetal:92d,ciuchini:92,ciuchinietal:93}.
Unfortunately, there exist sizable uncertainties in the hadronic
matrix elements of these operators which hopefully will be reduced
when new data are available. Moreover, strong cancellations between
QCD penguin and electroweak penguin contributions for large $m_t$ make
a precise theoretical prediction for $Re(\epsilon'/\epsilon)$ even
harder.  All efforts should be made to improve this situation.

        Whereas in $K \ra \pi \pi$ decays the CP violating
contribution is a tiny part of the full amplitude and the direct CP
violation is expected to be at least by three orders of magnitude
smaller than the indirect CP violation, the corresponding hierarchies
are very different for the rare decay \kpiee. At lowest order in
electroweak interactions (single photon, single Z-boson or double
W-boson exchange), this decay takes place only if CP symmetry is
violated \cite{gilman:79,gilman:80,dib1:89,dib2:89,flynn:89b}.
Moreover, the direct CP violating contribution might be larger than
the indirect one. The CP conserving contribution to the amplitude
comes from a two photon exchange,\footnote{Also the QED correction to
a single Z-boson or a W-box exchange can give CP conserving
contributions of the same order as the two photon exchange.} which
although higher order in $\alpha$ could be sizable.  The studies of
the last years \cite{cohenetal:93,pich:93} indicate however that the
CP conserving part is significantly smaller than the direct CP
violating contribution.

	The size of the indirect CP violating contribution will be
known once the CP conserving decay $K_S \ra \pi^0 e^+ e^-$ is
measured. On the other hand the direct CP violating contribution can
be fully calculated as a function of $m_t$, CKM parameters and the
QCD coupling constant $\al$. There are practically no theoretical
uncertainties related to hadronic matrix elements in this part,
because the latter can be extracted from the well-measured decay $K^+
\ra \pi^0 e^+ \nu$. In what follows, we will concentrate on this
contribution relegating the discussion of the other two contributions
to the end of the paper.

	The aim of this paper is to construct the effective
Hamiltonian for \kpiee,
\be \label{heff}
H_{eff} = \f{G_F}{\sqrt{2}} \left[ \sum_{i=1}^6 C_i(\mu) Q_i(\mu) +
C_{7V}(\mu) Q_{7V}(\mu) + C_{7A}(M_W) Q_{7A}(M_W) \right]
\ee
with the Wilson coefficients $C_i(\mu)$ including leading and
next-to-leading QCD corrections. Here $Q_{1,2}$ denote current-current
operators, $Q_{3\mbox{--}6}$ QCD penguin operators and
\be \label{7v7a}
Q_{7V} = (\bar{s} d)_{V-A} (\bar{e}e)_V,
\hspace{2cm}
Q_{7A} = (\bar{s} d)_{V-A} (\bar{e}e)_A
\ee
are the operators originating in the $\gamma$- and $Z^0$-penguin and
box diagrams of fig.~\ref{diag:2}.  For simplicity the CKM parameters
have been suppressed in (\ref{heff}). We will include them explicitly
in section \ref{sb.master}. Our paper can be considered as a
generalization of the analyses
\cite{gilman:79,gilman:80,dib1:89,dib2:89,flynn:89b} to include
next-to-leading logarithmic QCD effects. Whereas in
refs.~\cite{gilman:79,gilman:80,dib1:89,dib2:89,flynn:89b} the
logarithms $\alpha t (\al t)^n$ with $t=\ln(M^2_W/\mu^2)$ have been
summed, the present analysis includes also the summation of the
logarithms $\alpha (\al t)^n$ for which two-loop anomalous dimensions
of $Q_i$ are necessary. In this way we remove the sizeable
renormalization scheme dependence of the existing leading logarithmic
results. The numerical significance of such a renormalization scheme
dependence has been pointed out in ref.~\cite{misiak:91}.

        The important observation made in
refs.~\cite{dib1:89,dib2:89,flynn:89b} is the presence of
$Z^0$-penguin and box-diagram contributions which although small for
$m_t < M_W$, compete for $m_t > M_W$ with the photonic penguins. The
contributions of these new diagrams have been already included in
refs.~\cite{dib1:89,dib2:89,flynn:89b}. It should be stressed however
that these additional contributions are from the point of view of the
renormalization group at the next-to-leading order (NLO) and their
consistent inclusion requires also the summation of next-to-leading
logarithms. In view of the NLO analyses of
refs.~\cite{altarellietal:81,burasweisz:90,burasetal:92a,burasetal:92b,burasetal:92c,burasetal:92d,ciuchini:92,ciuchinietal:93},
the corresponding NLO analysis for \kpiee is now substantially easier
than it could have been at the time the calculations of
refs.~\cite{dib1:89,dib2:89,flynn:89b,misiak:91} were performed.

        With the complete calculation of next-to-leading contributions
we will be able to study for the first time the dependence of the
direct CP violation in \kpiee on the QCD scale parameter $\Lms$ and to
remove certain ambiguities present in the analyses of
refs.~\cite{dib1:89,dib2:89,flynn:89b,misiak:91}.

        Our paper is organized as follows. In section \ref{s.basic},
we present general formul\ae\ for $C_i(\mu)$ beyond the leading
logarithmic approximation and we discuss some aspects of the
renormalization scheme dependence. In section \ref{s.anom}, we
calculate the two-loop mixing between $Q_i\;\;\;(i=1,\ldots,6)$ and
$Q_{7V}$, which together with the two-loop results for $Q_i$ of
refs.~\cite{altarellietal:81,burasweisz:90,burasetal:92a,burasetal:92b,burasetal:92c,burasetal:92d,ciuchini:92,ciuchinietal:93}
gives the necessary two-loop anomalous dimension matrix for \kpiee. In
section \ref{s.ham}, we present the effective Hamiltonian relevant for
\kpiee and give numerical results for $C_i(\mu)$ in the NDR and HV
schemes. In section \ref{s.br}, we calculate $BR(K_L \ra \pi^0 e^+
e^-)_{dir}$ as a function of $m_t$ and $\Lms$ incorporating the
updated values of the CKM parameters. We also give an analytic
expression for $BR(K_L \ra \pi^0 e^+ e^-)_{dir}$ using the method of
Penguin-Box-Expansion \cite{buchallaetal:91}. In section
\ref{s.compar}, we compare the contribution of the direct CP violation
to \kpiee with the other two contributions. In section
\ref{s.sum} a brief summary and outlook are given. A few technical
details of the analysis have been relegated to the appendices.

\newsection{Basic Formul\ae\ for Wilson Coefficients} \label{s.basic}
\subsection{Operators} \label{sb.ope}

        Our basis of operators is given as follows
\be \label{oper}
\begin{array}{rcl}
Q_1    & = & (\bar{s}_{\alpha}  u_{\beta })_{V-A}
           (\bar{u}_{\beta }  d_{\alpha})_{V-A}    \vspace{0.2cm} \\
Q_2    & = & (\bar{s} u)_{V-A}  (\bar{u} d)_{V-A}    \vspace{0.2cm} \\
Q_3    & = & (\bar{s} d)_{V-A}\sum_q(\bar{q}q)_{V-A} \vspace{0.2cm} \\
Q_4    & = & (\bar{s}_{\alpha}  d_{\beta })_{V-A}
   \sum_q  (\bar{q}_{\beta }  q_{\alpha})_{V-A}    \vspace{0.2cm} \\
Q_5    & = & (\bar{s} d)_{V-A}\sum_q(\bar{q}q)_{V+A} \vspace{0.2cm} \\
Q_6    & = & (\bar{s}_{\alpha}  d_{\beta })_{V-A}
   \sum_q  (\bar{q}_{\beta }  q_{\alpha})_{V+A}    \vspace{0.2cm} \\
Q'_{7V}& = & (\alpha / \al)
           (\bar{s} d)_{V-A}  (\bar{e}e)_V         \vspace{0.2cm} \\
Q_{7A} & = & (\bar{s} d)_{V-A}  (\bar{e}e)_A         \vspace{0.2cm} \\
\end{array}
\ee
where $\alpha$ and $\beta$ denote colour indices
$(\alpha,\beta=1,\ldots,N)$. We omit the colour indices for the
colour-singlet currents. Labels $(V \pm A)$ refer to $\gamma_{\mu} (1 \pm
\gamma_5)$. The factor $\alpha/\al$ in the definition of $Q'_{7V}$
allows to make all the elements of the anomalous dimension matrix be
of the same order in $\al$. At the end of the renormalization group
analysis, this factor will be put back into the Wilson coefficient
$C_{7V}(\mu)$ of the operator $Q_{7V}$ in eq.~(\ref{7v7a}). There is
no need to include a similar factor in $Q_{7A}$ as this operator does
not mix under renormalization with the remaining operators. The sums
in eq.~(\ref{oper}) run over the flavours active at a given scale
$\mu$.

	For $\mu > m_c$ two additional current-current operators have
to be taken into account
\be \label{operc}
Q^c_1    = (\bar{s}_{\alpha}  c_{\beta })_{V-A}
           (\bar{c}_{\beta }  d_{\alpha})_{V-A}, \hspace{2cm}
Q^c_2    = (\bar{s} c)_{V-A}  (\bar{c} d)_{V-A}.
\ee

        It should be stressed that the form of $Q_1$, $Q_2$, $Q^c_1$ and
$Q^c_2$ here differs from the one used by Dib, Dunietz and Gilman
\cite{dib1:89,dib2:89} and by Flynn and Randall \cite{flynn:89b}
where the corresponding Fierz conjugates have been adopted
\be \label{tildeoperu}
\widetilde{Q}_1    = (\bar{s} d)_{V-A}  (\bar{u} u)_{V-A}, \hspace{2cm}
\widetilde{Q}_2    = (\bar{s}_{\alpha}  d_{\beta })_{V-A}
                  (\bar{u}_{\beta }  u_{\alpha})_{V-A},
\ee
\be \label{tildeoperc}
\widetilde{Q}^c_1    = (\bar{s} d)_{V-A}  (\bar{c} c)_{V-A}, \hspace{2cm}
\widetilde{Q}^c_2    = (\bar{s}_{\alpha}  d_{\beta })_{V-A}
                   (\bar{c}_{\beta }  c_{\alpha})_{V-A}.
\ee
Both bases are equally good although we somewhat prefer the basis
(\ref{oper}), because $Q_2$ is taken in the colour singlet form as it
appears in the tree level Hamiltonian. We will return later to the
comparison of next-to-leading order calculations performed in these
two bases.

        We do not include in our analysis the electroweak four-quark
penguin operators denoted as $Q_7$--$Q_{10}$ in
ref.~\cite{burasetal:92b} because their Wilson coefficients and matrix
elements for \kpiee are both of order ${\cal O}(\alpha)$ implying that
these operators enter the amplitude $A(K_L \ra \pi^0 e^+ e^-)$ at
${\cal O}(\alpha^2)$. This should be distinguished from the case of
$\epsilon'/\epsilon$. There, in spite of being suppressed by
$\alpha/\al$ relative to QCD penguin operators, the electroweak
penguin operators have to be included in the analysis because of the
additional enhancement factor $Re A_0/Re A_2 \simeq 22$. Such an
enhancement factor is not present here and the electroweak penguin
operators can be safely neglected.

        Concerning the Wilson coefficients, the electroweak four-quark
penguin operators affect through mixing under renormalization the
coefficients $C_3$--$C_6$ at ${\cal O}(\alpha)$ and $C_{7V}$ at ${\cal
O}(\alpha^2)$. Since the corresponding matrix elements are ${\cal
O}(\alpha)$ and ${\cal O}(1)$ respectively, we again obtain a
negligible ${\cal O}(\alpha^2)$ effect in $A(K_L \ra \pi^0 e^+ e^-)$.

        We also neglect the ``magnetic moment'' operators which play a
crucial role in the $b \ra s \gamma$ and $s \ra d \gamma$ transitions.
These operators, being of dimension five, do not influence the Wilson
coefficients of the operators in eq.~(\ref{oper}). Since their
contributions to \kpiee are suppressed by an additional factor $m_s$,
they appear strictly speaking at higher order in chiral perturbation
theory.

\subsection{Renormalization Group Equations} \label{sb.RGE}

        The renormalization group equation for $\vec{C}(\mu)$ is given
by
\be \label{RGE}
\left[\mu \f{\partial}{\partial \mu} + \beta(g) \f{\partial}{\partial
g} \right] \vec{C}(\f{M^2_W}{\mu^2},g) = \hat{\gamma}(g)
\vec{C}(\f{M^2_W}{\mu^2},g)
\ee
where $\beta(g)$ is the QCD beta function
\be \label{expbet}
\beta(g) = -\beta_0 \f{g^3}{16 \pi^2} - \beta_1 \f{g^5}{(16 \pi^2)^2}-
\ldots,
\ee
with
\be
\beta_0=11-\f{2}{3}f, \hspace{2cm} \beta_1=102-\f{38}{3}f,
\ee
and $f=u+d$ denoting the number of active flavours, $u$ and $d$ being the
number of $u$-type and $d$-type flavours, respectively.

        In what follows, we will neglect the running of the
electromagnetic coupling constant $\alpha$. This is a very good
approximation because only scales $1 \gev \leq \mu \leq M_W$ are
involved in our analysis. In the numerical analysis we will take
$\alpha = 1/128$. For the effective QCD coupling constant we will use
\be
\label{alpha}
\al^{(f)}(Q) = \f{4 \pi}{\beta_0 \ln(Q^2/\Lambda_f^2)} \left[ 1 -
\f{\beta_1 \ln\;\ln(Q^2/\Lambda_f^2)}{\beta_0^2 \ln(Q^2/\Lambda_f^2)} \right],
\ee
with $\Lambda_f$ in the $\overline{MS}$ scheme. Demanding the
continuity of $\al^{(f)}(Q)$ at quark thresholds in the form
\be
\al^{(3)}(m_c) = \al^{(4)}(m_c), \hspace{2cm} \al^{(4)}(m_b) = \al^{(5)}(m_b)
\ee
gives relations between the various $\Lambda_f$. In what follows we
will denote $\Lms \equiv \Lambda_4 \equiv \Lambda_{QCD}$.

        Although it has recently become popular to use $\al(M_Z)$
instead of $\Lambda_{QCD}$ to parametrize the experimental knowledge
about $\al$, we will present our results for the Wilson coefficients
as functions of $\Lambda_4$ in order to facilitate the comparison with
refs.~\cite{dib1:89,dib2:89,flynn:89b}. For each particular value of
$\Lambda_4$ used, we give the corresponding value of $\al(M_Z)$ with
the next-to-leading order accuracy in app.~\ref{parameters}.

        Since the operator $Q_{7A}$ does not mix with the remaining
operators in eq.~(\ref{oper}), we will not include its coefficient in
$\vec{C}(\mu)$. Consequently, $\hat{\gamma}(g)$ in eq.~(\ref{RGE}) is a
$7 \times 7$ anomalous dimension matrix which we expand as follows
\be \label{expgam}
\hat{\gamma}(g) = \hat{\gamma}^{(0)} \f{g^2}{16 \pi^2} +
                  \hat{\gamma}^{(1)} \f{g^4}{(16 \pi^2)^2} +  \ldots\, .
\ee
The calculation of this matrix will be presented in section \ref{s.anom}.

\subsection{Solution for $\vec{C}(\mu)$} \label{sb.solution}

        Following \cite{burasetal:92a} we write the solution of
eq.~(\ref{RGE}) as
\be
\vec{C}(\mu) = \hat{U}(\mu,M_W) \vec{C}(M_W)
\ee
where the renormalization group evolution matrix is given generally by
\be \label{umatrix1}
\hat{U}(m_1,m_2) = T_g \exp \int_{g(m_2)}^{g(m_1)} dg'
\f{\hat{\gamma}^T(g')}{\beta(g')}
\ee
with $m_1 < m_2$. Here $T_g$ denotes such ordering in the coupling
constants that they increase from right to left. Expanding
$\hat{\gamma}(g)$ and $\beta(g)$ as in eqs.~(\ref{expgam}) and
(\ref{expbet}) respectively one finds
\be \label{umatrix}
\hat{U}(m_1,m_2) = \left( \hat{1} + \f{\al(m_1)}{4 \pi} \hat{J} \right)
                        \hat{U}^{(0)}(m_1,m_2) \left( \hat{1} -
                   \f{\al(m_2)}{4 \pi} \hat{J} \right)
\ee
where $\hat{U}^{(0)}(m_1,m_2)$ denotes the evolution matrix
in the leading logarithmic approximation and $\hat{J}$ summarizes the
next-to-leading corrections to this evolution. If
\be
\hat{\gamma}^{(0)}_D \equiv \hat{V}^{-1}\hat{\gamma}^{(0)T}\hat{V},
\hspace{3cm}
\hat{G} \equiv \hat{V}^{-1}\hat{\gamma}^{(1)T} \hat{V}
\ee
where $\hat{\gamma}^{(0)}_D$ denotes a diagonal matrix whose
diagonal elements are the components of the vector
$\vec{\gamma}^{(0)}$, then
\be
\hat{U}^{(0)}(m_1,m_2) =
\hat{V}
\left[ \left( \f{\al(m_2)}{\al(m_1)} \right)^{\vec{a}} \right]_D
\hat{V}^{-1}
\hspace{0.5cm} {\rm with} \hspace{0.5cm}
\vec{a} = \f{\vec{\gamma}^{(0)}}{2 \beta_0}.
\ee
For the matrix $\hat{J}$ one gets
\be
\hat{J} = \hat{V} \hat{S} \hat{V}^{-1}
\ee
where the elements of $\hat{S}$ are given by
\be
S_{ij} = \delta_{ij} \gamma_i^{(0)} \f{\beta_1}{2 \beta_0^2} -
\f{G_{ij}}{2 \beta_0 + \gamma_i^{(0)} - \gamma_j^{(0)}}
\ee
with $\gamma_i^{(0)}$ denoting the elements of $\vec{\gamma}^{(0)}$
and $G_{ij}$ the elements of $\hat{G}$.

\subsection{The Initial Conditions $\vec{C}(M_W)$} \label{sb.match}

        The calculation of the initial conditions $\vec{C}(M_W)$ has
been discussed at length in
refs.~\cite{dib1:89,dib2:89,flynn:89b,burasetal:92d,ciuchini:92}. In
fact, the results for $\vec{C}(M_W)$ can be extracted from these
papers. Therefore we recall here only the basic ingredients of this
calculation stressing one point which has not been discussed in
previous analyses of \kpiee.

\figintext{
\begin{figure}[h]
\vspace{0.10in}
\centerline{
\epsfysize=3.5in
\epsffile{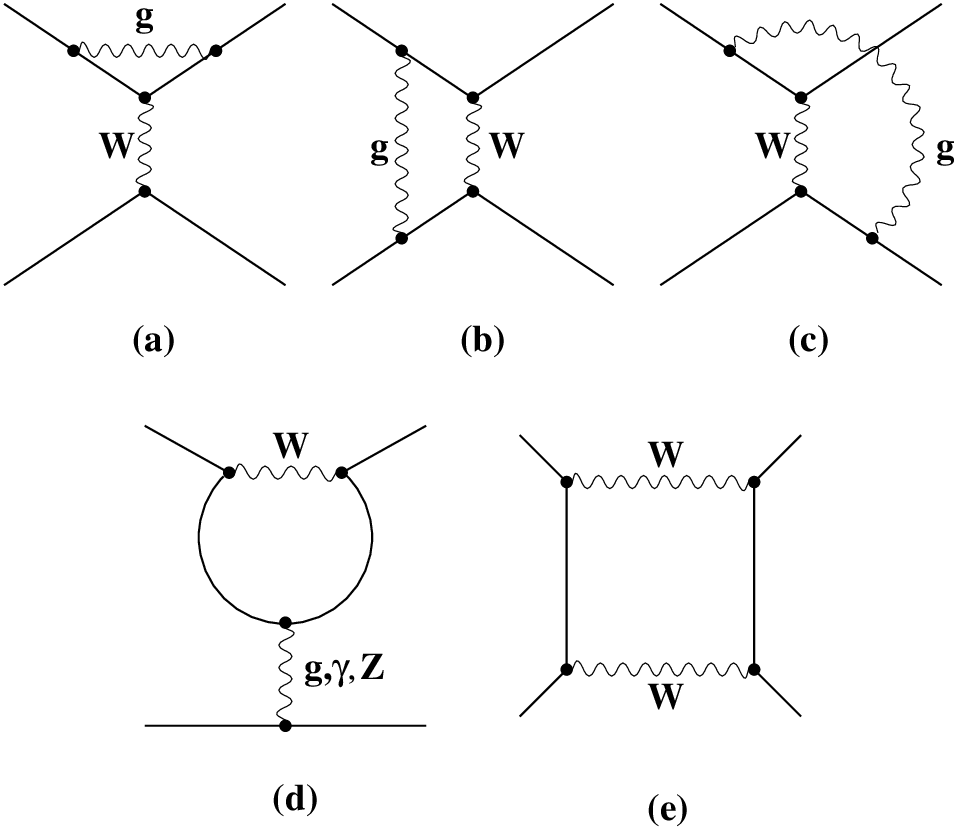}
}
\vspace{0.08in}
\caption[]{
One-loop current-current and penguin diagrams in the full theory.
\label{diag:2}}
\end{figure}
}

\figintext{
\begin{figure}[hbt]
\vspace{0.10in}
\centerline{
\epsfysize=2in
\epsffile{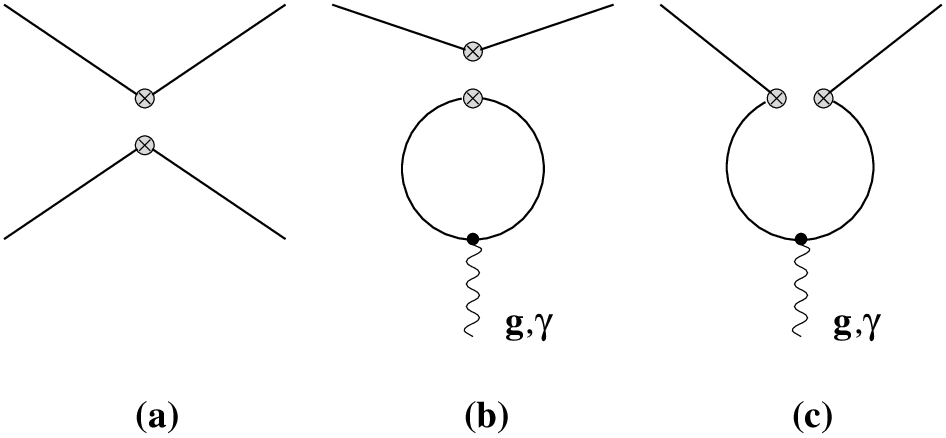}
}
\vspace{0.08in}
\caption[]{
The three basic ways of inserting a given operator into a four-point
function: (a) current-current-, (b) type-1 penguin-, (c) type-2
penguin-insertion. The 4-vertices ``$\otimes\ \otimes$'' denote
standard operator insertions.
\label{diag:1}}
\end{figure}
}

        In order to find $\vec{C}(M_W)$, the one-loop current-current
and penguin diagrams of fig.~\ref{diag:2} with the full W and Z
propagators and internal top quark exchanges have to be calculated
first. Also the relevant counterterms have to be added to make the
result finite. Subsequently, the result of this calculation has to be
expressed in terms of matrix elements $\me{\vec{Q}(M_W)}$. The latter
are found by inserting the operators $\vec{Q}$ as depicted in
fig.~\ref{diag:1} into the one-loop current-current and penguin
diagrams of fig.~\ref{diag:3} and calculating the finite contributions
in some renormalization scheme.

\figintext{
\begin{figure}[htb]
\vspace{0.10in}
\centerline{
\epsfysize=3.5in
\epsffile{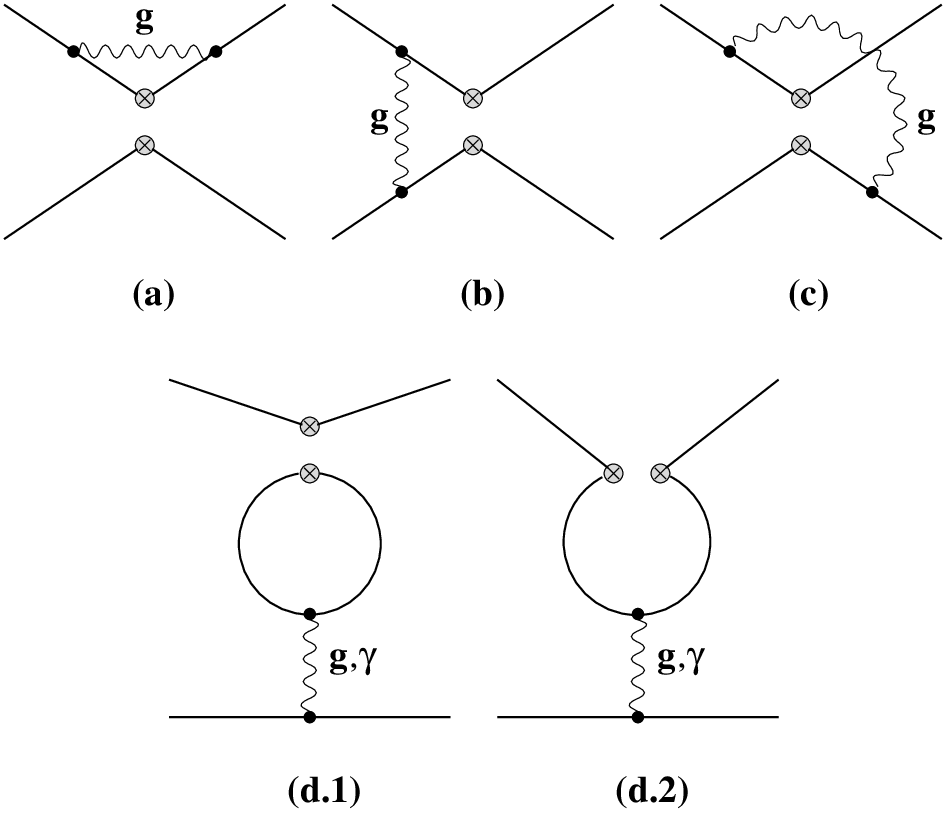}
}
\vspace{0.08in}
\caption[]{
One-loop current-current and penguin diagrams contributing to
$\gamma^{(0)}$ and matching conditions in the effective theory. The
meaning of vertices is the same as in fig~\ref{diag:1}. Possible
left-right or up-down reflected diagrams are not shown.
\label{diag:3}}
\end{figure}
}

        The resulting coefficients $\vec{C}(M_W)$ do not depend on the
assumptions made about the properties of the external lines in
figs.~\ref{diag:2} and \ref{diag:3}, i.e. on the infrared structure of
the theory. They depend however on the renormalization scheme through
the matrix elements $\me{\vec{Q}(M_W)}$. In our calculation we will
use as in
\cite{burasweisz:90,burasetal:92a,burasetal:92b,burasetal:92c,burasetal:92d,ciuchini:92,ciuchinietal:93}
two renormalization schemes usually called NDR and HV.

        For the NDR scheme and the basis (\ref{oper}) one finds
\be \label{c1c2}
C_1(M_W) = \f{11}{2} \f{\al(M_W)}{4 \pi},
\hspace{2cm}
C_2(M_W) = 1-\f{11}{6} \f{\al(M_W)}{4 \pi},
\ee
\be \label{c3-c6}
C_3(M_W) = -\f{1}{3} C_4(M_W) = C_5(M_W) = -\f{1}{3} C_6(M_W)
= -\f{\al(M_W)}{24 \pi} \widetilde{E}(x_t),
\ee
\be \label{c7v}
C'_{7V}(M_W) = \f{\al(M_W)}{2 \pi}
\left[ \f{C(x_t)-B(x_t)}{\sin^2\theta_W} -\widetilde{D}(x_t) - 4 C(x_t) \right]
\ee
and
\be \label{c7a}
C_{7A}(M_W) = \f{\alpha}{2 \pi} \f{B(x_t)-C(x_t)}{\sin^2\theta_W}.
\ee
Here
\be \label{tildefunc}
\widetilde{E}(x_t) = E(x_t) - \f{2}{3},
\hspace{2cm}
\widetilde{D}(x_t) = D(x_t) - \f{4}{9},
\ee
\be
x_t = \f{m_t^2}{M_W^2} \, ,
\ee
and
\begin{eqnarray}
\label{BCDE}
B(x) &=& \f{1}{4} \left[ \f{x}{1-x} + \f{x \ln x}{(x-1)^2} \right],\\
C(x) &=& \f{x}{8} \left[ \f{x-6}{x-1} + \f{3 x + 2}{(x-1)^2} \ln x
\right],\\
D(x) &=& \f{-19 x^3 + 25 x^2}{36 (x-1)^3}
+ \f{x^2 (5 x^2 - 2 x - 6)}{18 (x-1)^4} \ln x- \f{4}{9} \ln x,\\
E(x) &=& \f{x (18 -11
x - x^2)}{12 (1-x)^3} + \f{x^2 (15 - 16 x + 4 x^2)}{6 (1-x)^4} \ln
x-\f{2}{3} \ln x.
\end{eqnarray}
Comparing these results with the initial conditions used in
refs.~\cite{dib1:89,dib2:89,flynn:89b} we notice some differences. In
refs.~\cite{dib1:89,dib2:89,flynn:89b} the leading order
evolution matrices have been used but in the initial conditions the
$m_t$-dependent next-to-leading terms have been already taken into
account. These $m_t$-dependent terms represented by the functions
$B(x_t)$, $C(x_t)$, $D(x_t)$ and $E(x_t)$ are renormalization scheme
independent. Yet, the initial conditions may also contain scheme
dependent next-to-leading terms which of course were not present in
the analyses of refs.~\cite{dib1:89,dib2:89,flynn:89b}. In the NDR
scheme these scheme dependent terms enter the initial conditions for
$Q_1$ and $Q_2$ and also modify the functions $E(x_t)$ and $D(x_t)$ by
constant terms as given in eq.~(\ref{tildefunc}).

        The additional constant terms in eq.~(\ref{tildefunc}) are
absent in the case of the basis (\ref{tildeoperu})--(\ref{tildeoperc})
used in refs.~\cite{dib1:89,dib2:89,flynn:89b}. However, as pointed
out in ref.~\cite{burasetal:92a,burasetal:92d}, in this case also the
corresponding two-loop anomalous dimensions $\hat{\gamma}^{(1)}$
change so that the physical amplitudes remain independent of whether
the basis (\ref{oper}) or (\ref{tildeoperu})--(\ref{tildeoperc}) is
used. This clarifies the observation made by Flynn and Randall
\cite{flynn:89b} that the initial conditions for penguin operators do
depend on the form of the operators. This dependence signals the
necessity of going beyond the leading logarithmic approximation for
$\hat{U}(m_1,m_2)$ in order to obtain physical results.

        In the HV scheme the constant terms in eq.~(\ref{tildefunc})
are also absent. In addition the initial conditions $C_{1,2}(M_W)$
change. They can be found in app.~\ref{HV}. Again, in physical
quantities this change is cancelled by the corresponding change in the
two-loop anomalous dimension matrix $\hat{\gamma}^{(1)}$ calculated in
the HV scheme.

\subsection{Renormalization Scheme Dependence of $\hat{\gamma}^{(1)}$}
\label{sb.rendep}

        The renormalization scheme dependence of $\hat{\gamma}^{(1)}$
and the cancellation of scheme dependences in the final physical
amplitudes have been discussed at length in
refs.~\cite{burasweisz:90,burasetal:92a,burasetal:92b,burasetal:92c,burasetal:92d,ciuchini:92,ciuchinietal:93}.
Here we would like only to recall one important relation between
$(\hat{\gamma}^{(1)})_a$ and $(\hat{\gamma}^{(1)})_b$ calculated in
two different renormalization schemes $a$ and $b$ or two different
bases like (\ref{oper}) and (\ref{tildeoperu})--(\ref{tildeoperc}). We
have
\be \label{rendep}
(\hat{\gamma}^{(1)})_b = (\hat{\gamma}^{(1)})_a + \left[ \Delta
\hat{r}, \hat{\gamma}^{(0)} \right]  + 2 \beta_0 \Delta \hat{r}
\ee
with $\Delta \hat{r} \equiv (\hat{r})_b - (\hat{r})_a$, where
$(\hat{r})_i$ are defined by
\be
\label{me}
\me{\vec{Q}(M_W)}_i = \left[ \hat{1} + \f{\al(M_W)}{4 \pi} (\hat{r})_i
\right] \me{\vec{Q}^{(0)}}.
\ee

        Here, $\vec{Q}^{(0)}$ is a tree-level matrix element and
$\me{\vec{Q}}_i$ denotes renormalized one-loop matrix elements
calculated in the scheme $i$. The matrices $(\hat{r})_i$ are obtained
by calculating the finite terms in the one-loop diagrams of
fig.~\ref{diag:3}.

        Relation (\ref{rendep}) is very useful as it allows to test
compatibility of the two-loop anomalous dimension matrices calculated
in two different renormalization or regularization schemes. It also
plays a central role in demonstrating the scheme independence of
physical quantities \cite{burasetal:92a,burasetal:92d}. Moreover, as
demonstrated in refs.~\cite{burasetal:92b,burasetal:92c} and in the
app.~\ref{bases} here, the relation (\ref{rendep}) allows to avoid
explicit calculation of two-loop diagrams involving traces
$Tr(\gamma_{\mu}\gamma_{\nu}\gamma_{\rho}\gamma_{\lambda}\gamma_5)$
which are dangerous in the NDR scheme.

\newsection{Anomalous Dimension Matrices} \label{s.anom}
\subsection{The Matrix $\hat{\gamma}^{(0)}$} \label{sb.gamma0}

        The $6 \times 6$ submatrix of $\hat{\gamma}^{(0)}$ involving
the operators $Q_1$--$Q_6$ has been calculated in
ref.~\cite{gilman:79,gaillardlee:74,altarellimaiani:74,SVZ:77,guberinapecchei:80}.  In
our normalizations it is given in app.~A of
ref.~\cite{burasetal:92b}. Here we only give the remaining non-vanishing
entries of $\hat{\gamma}^{(0)}$:
\begin{eqnarray}
\label{17,27:LO}
\gamma^{(0)}_{17} & = & -\f{16}{9} N \hspace{3.8cm} \gamma^{(0)}_{27} =
-\f{16}{9} \\
\gamma^{(0)}_{37} & = & -\f{16}{9} N \left( u -\f{d}{2} - \f{1}{N} \right)
\hspace{1cm} \gamma^{(0)}_{47}  = -\f{16}{9} \left( u -\f{d}{2} - N
\right) \\
\label{57,67:LO}
\gamma^{(0)}_{57} & = & -\f{16}{9} N \left( u -\f{d}{2} \right) \hspace{2cm}
\gamma^{(0)}_{67} = -\f{16}{9} \left( u -\f{d}{2} \right) \\
\gamma^{(0)}_{77} & = & -2 \beta_0 = -\f{22}{3} N + \f{4}{3} f
\end{eqnarray}
where N is the number of colours. These elements have been first
calculated in \cite{gilman:80} except that $\gamma^{(0)}_{37}$ and
$\gamma^{(0)}_{47}$ have been corrected in
refs.~\cite{flynn:89b,eegpicek:88}.  We agree with these final
results.

\subsection{The Matrix $\hat{\gamma}^{(1)}$} \label{sb.gamma1}

        The $6 \times 6$ submatrix of $\hat{\gamma}^{(1)}$ involving
the operators $Q_1$--$Q_6$ has been calculated in
ref.~\cite{burasetal:92a,ciuchinietal:93} in the NDR and HV
renormalization schemes. The results presented there include also the
ones obtained in refs.~\cite{altarellietal:81,burasweisz:90} where the
two-loop mixing of $Q_1$ and $Q_2$ operators has been considered. In
our normalizations, the submatrix in question is given in app.~B
of ref.~\cite{burasetal:92b}.

        In order to calculate the remaining entries of
$\hat{\gamma}^{(1)}$, one has to evaluate the two-loop penguin
diagrams of fig.~\ref{diag:5}. Since these diagrams constitute a
subset of the diagrams considered in \cite{burasetal:92b}, all the
singularities necessary for the evaluation of the entries in question
can be extracted for instance from tab.~2 of
ref.~\cite{burasetal:92b}.  Consequently, it was enough to change the
relevant colour and electric charge factors in the diagrams of
fig.~\ref{diag:5}.

\figintext{
\begin{figure}[htb]
\vspace{0.10in}
\centerline{
\epsfysize=4in
\epsffile{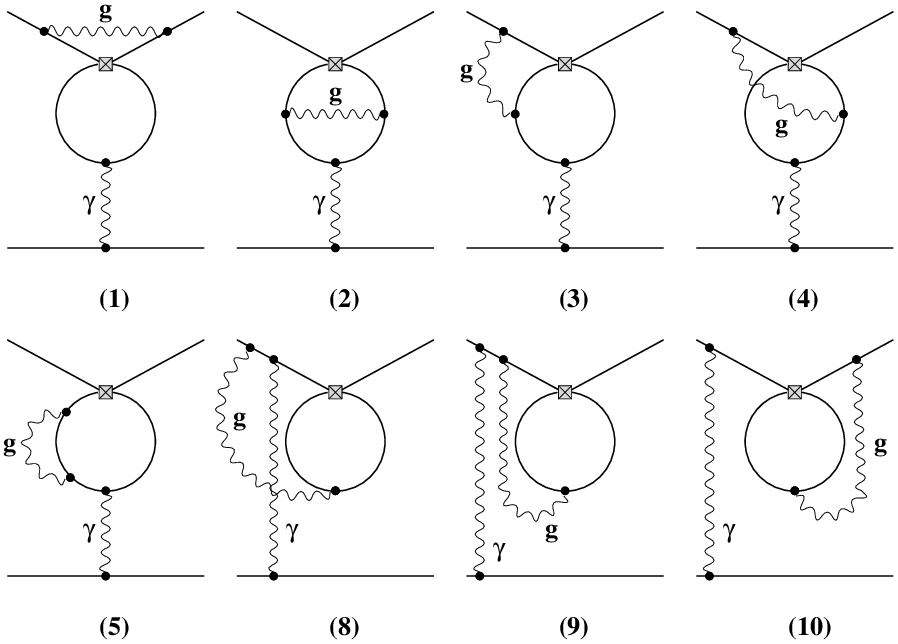}
}
\vspace{0.08in}
\caption[]{
Two--loop penguin diagrams contributing to $\gamma_{i7}^{(1)}$,
$i=1,\ldots,6$. Square-vertices stand for type-1 and type-2 penguin
insertions as of figs.~\ref{diag:1}(b) and (c), respectively.
Possible left-right reflected diagrams are not shown.
\label{diag:5}}
\end{figure}
}

        Our final results for the seventh column of
$\hat{\gamma}^{(1)}$ are given in the NDR scheme as follows:\footnote
{The first six entries in the seventh row of $\hat{\gamma}^{(1)}$ vanish.}
\begin{eqnarray}
\label{17,27}
\gamma^{(1)}_{17} & = & \f{8}{3} \left( 1 - N^2 \right), \hspace{1.5cm}
\gamma^{(1)}_{27} = \f{200}{81} \left( N - \f{1}{N} \right), \\
\label{37}
\gamma^{(1)}_{37} & = & \f{8}{3} \left( u - \f{d}{2} \right) \left( 1
- N^2 \right) + \f{464}{81} \left( \f{1}{N} - N \right), \\
\label{47}
\gamma^{(1)}_{47} & = & \left( u \f{280}{81} + d \f{64}{81} \right)
\left( \f{1}{N} - N \right) + \f{8}{3} \left( N^2 - 1 \right), \\
\label{57}
\gamma^{(1)}_{57} & = & \f{8}{3} \left( u - \f{d}{2} \right) \left(
1 - N^2 \right), \\
\label{67}
\gamma^{(1)}_{67} & = & \left( u \f{440}{81} - d \f{424}{81} \right)
\left( N - \f{1}{N} \right), \\
\label{77}
\gamma^{(1)}_{77} & = & -2 \beta_1 = -\f{68}{3} N^2 + \f{20}{3} N f + 4 C_F f
\end{eqnarray}
where $C_F = \f{N^2 -1}{2N}$. The corresponding results for the HV
scheme can either be obtained by direct calculation or by using the
relation (\ref{rendep}). They can be found in app.~\ref{HV}.

	As shown in app.~\ref{mikolaj} of the present paper,
the results (\ref{17,27})--(\ref{77}) are in agreement with those
calculated independently in ref.~\cite{misiak:93}, as long as the
differences in the conventions used are properly taken into account.
However, the present paper is the first one where the the two-loop
mixing among $Q_1$--$Q_6$ from ref.~\cite{burasetal:92a} is
consistently put together with the mixing of $Q_1$--$Q_6$ into
$Q_{7V}$. The author of ref.~\cite{misiak:93} did not take into
account that certain conventions used in ref.~\cite{burasetal:92a}
were different from his own, and did not transform the results of
ref.~\cite{burasetal:92a} into his own conventions (see
app.~\ref{mikolaj} for details).

	The calculation of two-loop mixings made in the NDR scheme
requires some care when
$Tr(\gamma_{\mu}\gamma_{\nu}\gamma_{\rho}\gamma_{\lambda}\gamma_5)$
appears in the calculation. One of the ways dealing with this problem
is just to avoid appearance of such traces by making use of the fact
that a Fierz rearrangement of any of the considered operators should
not affect physical results. This method has been discussed at length
in refs.~\cite{burasetal:92b,burasetal:92c}. It is straightforward to
extend it to the case of the operator basis containing also $Q_{7V}$.
For the more complex cases of $\gamma^{(1)}_{37}$ and
$\gamma^{(1)}_{47}$ we describe this extension in
app.~\ref{bases}. This is the method we have used to obtain the
results (\ref{37}) and (\ref{47}).

        Another method described in app.~A of ref.~\cite{misiak:93}
amounts to treating fermion lines with single
$\left\{\gamma_{\mu},\gamma_5\right\}$ as evanescent structures, i.e.
structures that vanish in 4 dimensions but have to be treated as
independent basic structures in D dimensions. In consequence, no
algebraic inconsistencies with anticommuting $\gamma_5$
arise.\footnote{The triangle anomaly is also correctly reproduced.}
Getting rid of the trace
$Tr(\gamma_{\mu}\gamma_{\nu}\gamma_{\rho}\gamma_{\lambda}\gamma_5)$ is
achieved by introducing other evanescent structures. In the particular
calculation of the two-loop mixing among the operators of the basis
(\ref{oper}), the new evanescent structures can be shown to give no
contribution. So the calculation in our particular case effectively
amounts to anticommuting $\gamma_5$ whenever necessary (except for traces
with single $\gamma_5$) and using eq.~(19) of ref.~\cite{misiak:93} (or,
equivalently, eq.~(5.20) of ref.~\cite{burasetal:92b}) to get rid of the
traces with $\gamma_5$.

        Actually, in
refs.~\cite{burasetal:92a,burasetal:92b,burasetal:92c} and in
ref.~\cite{misiak:93} the two approaches were used and checked to
give the same results. Also the equivalence of the physical results
to the ones obtained within the HV scheme has been checked in
refs.~\cite{burasetal:92a,burasetal:92b,burasetal:92c,ciuchini:92,ciuchinietal:93}
for the case of mixing among the four-quark operators.

\newsection{The Effective Hamiltonian for \kpiee\hspace{1cm} Beyond Leading
Logarithms} \label{s.ham}
\subsection{Master Formul\ae\ for Wilson Coefficient Functions}
\label{sb.master}

        Integrating out simultaneously $W$, $Z$ and $t$ we construct
the effective Hamiltonian for $\Delta S =1$ transitions relevant for
\kpiee with the operators normalized at $\mu=M_W$. Following the
procedure of section \ref{sb.match} we find
\be \label{tausplit}
H_{eff}(\Delta S =1) = \f{G_F}{\sqrt{2}} V_{ud} V^*_{us} \left\{
(1-\tau) H^{uc}(M_W) + \tau H^{ut}(M_W) \right\}
\ee
where
\begin{eqnarray}
\label{cw}
H^{uc}(M_W) & = & C_1(M_W)(Q_1-Q_1^c) + C_2(M_W)(Q_2-Q_2^c), \\
\label{tw}
H^{ut}(M_W) & = & \sum_{i=1}^6 C_i(M_W) Q_i + C'_{7V}(M_W) Q'_{7V} +
C_{7A}(M_W) Q_{7A}
\end{eqnarray}
and
\be
\tau = - \f{V_{td} V^*_{ts}}{V_{ud} V^*_{us}}.
\ee
Here $Q_i$ and $Q^c_{1,2}$ are the operators defined in
eqs.~(\ref{oper})--(\ref{operc}) with $q=u,d,s,c,b$. Due to the GIM mechanism,
there are no contributions of $Q_i(i \neq 1,2)$ to $H^{uc}(M_W)$. The
coefficients $C_i(M_W)$, $C'_{7V}(M_W)$ and $C_{7A}(M_W)$ are given in
eqs.~(\ref{c1c2})--(\ref{c7a}).

        We next integrate out successively the $b$- and $c$-quarks and
transform $H_{eff}$ of eq.~(\ref{tausplit}) from $\mu=M_W$ down to $\mu <
m_c$ relevant for \kpiee. This transformation can be compactly
described as follows.
\begin{eqnarray}
\label{cmu}
H^{uc}(M_W) & \longrightarrow & \sum_{i=1}^6 z_i(\mu) Q_i + z'_{7V}(\mu)
Q'_{7V}, \\
\label{tmu}
H^{ut}(M_W) & \longrightarrow & \sum_{i=1}^6 v_i(\mu) Q_i +
v'_{7V}(\mu) Q'_{7V}+ C_{7A}(M_W) Q_{7A}
\end{eqnarray}
with the operators $Q_i$ and $Q'_{7V}$ renormalized at the scale $\mu$.
There is no renormalization of $Q_{7A}$. Since the charm quark has
been integrated out, the penguin operators contribute also in eq.~(\ref{cmu}).

        In writing eqs.~(\ref{cmu}) and (\ref{tmu}) we have neglected
the internal $u$- and $c$-quark contributions to the coefficient of the
operator $Q_{7A}$ because due to the vanishing of $C(x)$ and $B(x)$ as
$x\, \ln x$ for $x \ra 0$ they are very small compared to the top
contribution.

        The coefficients $(z_i,z'_{7V})$ and $(v_i,v'_{7V})$ are
components of the seven-dimensi\/onal column vectors
\be
\vec{z}(\mu) = \hat{U}_3(\mu,m_c) \vec{z}(m_c)
\ee
and
\be
\vec{v}(\mu) = \hat{U}_3(\mu,m_c) \hat{M}(m_c)\hat{U}_4(m_c,m_b)
\hat{M}(m_b)\hat{U}_5(m_b,M_W) \vec{C}(M_W)
\ee
with the components of $\vec{C}(M_W)$ given in eqs.~(\ref{c1c2})--(\ref{c7v}).

        The evolution matrices $U_f$ are given by eq.~(\ref{umatrix})
with $f$ denoting the number of active flavours. $\hat{M}(m_i)$ is
the matching matrix at quark threshold $m_i$. It will be discussed
separately in section \ref{sb.hatM}.

        Inserting eqs.~(\ref{cmu}) and (\ref{tmu}) into eq.~(\ref{tausplit})
and introducing the coefficients
\be
v_{7V}(\mu) = \f{\alpha}{\al(\mu)}\; v'_{7V}(\mu),
\hspace{2cm}
z_{7V}(\mu) = \f{\alpha}{\al(\mu)}\; z'_{7V}(\mu)
\ee
we find the effective Hamiltonian relevant for the decay \kpiee
\be
H_{eff} = \f{G_F}{\sqrt{2}} V_{ud} V^*_{us} \left[ \sum_{i=1}^{7V}
(z_i(\mu) + \tau y_i(\mu)) Q_i(\mu) + \tau y_{7A}(M_W) Q_{7A} \right]
\ee
where
\be
y_i(\mu) = v_i(\mu) - z_i(\mu), \hspace{3cm} y_{7A}(M_W) = C_{7A}(M_W).
\ee
In order to find $\vec{z}(\mu)$ we still need $\vec{z}(m_c)$. Due to
the GIM mechanism, the coefficients $z_i(\mu)$ of the penguin operators
$Q_i$, $i \neq 1,2$ are zero in 5- and 4-flavour theories. The
evolution in this range of $\mu$ ($\mu > m_c$) involves then only
current-current operators $Q_{1,2} \equiv Q^u_{1,2}$~and~$Q^c_{1,2}$
with the initial conditions
\be
z_1(M_W) = C_1(M_W), \hspace{2cm} z_2(M_W) = C_2(M_W)
\ee
where $C_{1,2}(M_W)$ are given in eq.~(\ref{c1c2}). $Q^u_{1,2}$ and
$Q^c_{1,2}$ do not mix under renormalization, and the coefficients of
these two pairs of operators are equal for $\mu \geq m_c$. We then find
\be \label{2*2evol}
\left( \begin{array}{c} z_1(m_c) \\ z_2(m_c) \end{array} \right)
= \hat{U}_4(m_c,m_b) \hat{M}(m_b)\hat{U}_5(m_b,M_W)
\left( \begin{array}{c} z_1(M_W) \\ z_2(M_W) \end{array} \right)
\ee
where this time the evolution matrices $\hat{U}_{4,5}$ contain only
the $2\times 2$ anomalous dimension matrices describing the mixing
between current-current operators. The matching matrix $\hat{M}(m_b)$
is in this case a $2\times 2$ unit matrix. When the charm quark is
integrated out, the operators $Q^c_{1,2}$ disappear from the effective
Hamiltonian and the coefficients $z_i(\mu),\;i \neq 1,2$ for penguin
operators become non-zero. In order to calculate $z_i(m_c)$ for
penguin operators, a proper matching between effective 4- and 3-quark
theories has to be made. This matching has been already discussed in
detail in section 4 of ref.~\cite{burasetal:92d} where the first six
components of $\vec{z}(m_c)$ are given. In order to find
$z'_{7V}(m_c)$ which results from the diagrams of fig.~\ref{diag:3},
we simply have to rescale the result for the $z_7(m_c)$ in eq.~(4.29)
of ref.~\cite{burasetal:92d} by a factor of $-3 \al/\alpha$.

        In summary, the components of $\vec{z}(m_c)$ are given by
eq.~(\ref{2*2evol}),
\be
z_3(m_c) = -\f{z_4(m_c)}{3} = z_5(m_c) = -\f{z_6(m_c)}{3} =
-\f{\al(m_c)}{24\pi} F_s(m_c)
\ee
and
\be
z'_{7V}(m_c) = -\f{\al(m_c)}{2 \pi} F_e(m_c)
\ee
where
\begin{eqnarray}
\label{Fs}
F_s(m_c) & = & \left\{ \begin{array}{ll} -\f{2}{3} z_2(m_c), & NDR \\
0, & HV. \end{array}\right. \\
\label{Fe}
F_e(m_c) & = & \left\{ \begin{array}{ll} -\f{4}{9} (3z_1(m_c)+z_2(m_c)), &
NDR \\ 0, & HV. \end{array} \right.
\end{eqnarray}
In the NDR scheme these results are given for the basis (\ref{oper}).

\subsection{The Matching Matrix $\hat{M}$} \label{sb.hatM}

        The additional matching matrix $\hat{M}$ between the operators
in an $f$-flavour effective theory and an $(f+1)$-flavour effective
theory is necessary because the QCD penguin operators have different
structure in different effective theories. In our case
\be
\hat{M}(m) = \hat{1} + \f{\al(m)}{4 \pi} \delta \hat{r}_s^T
\ee
where $\delta \hat{r}_s^T$ is a $7 \times 7$ matrix.

        The procedure for finding $\hat{M}(m)$ has been outlined in
ref.~\cite{burasetal:92d} where the $6 \times 6$ submatrix involving
$Q_1$--$Q_6$ has been calculated. The remaining entries of $\delta
\hat{r}_s$ can be found from the matrix $\delta \hat{r}_e$, also
calculated in ref.~\cite{burasetal:92d}, by making a simple rescaling by $-3
\al/\alpha$ as in the case of $\vec{z}(m_c)$.

        In summary, $\delta \hat{r}_s$ is given as follows. The
non-vanishing rows in the $6 \times 6$ submatrix are given for
$\mu=m_b$ and $\mu=m_c$ by
\be
(\delta \hat{r}_s)_{4i} = (\delta \hat{r}_s)_{6i} = -\f{5}{9} P_i
\ee
where
\be
\vec{P} = (0,0,-\f{1}{3},1,-\f{1}{3},1).
\ee
The seventh row of $\delta \hat{r}_s$ vanishes. The seventh column has
the following non-vanishing elements for $\mu=m_b$:
\be
(\delta \hat{r}_s)_{37} = 3 (\delta \hat{r}_s)_{47} = (\delta
\hat{r}_s)_{57} = 3 (\delta \hat{r}_s)_{67} = -\f{20}{9}.
\ee
For $\mu=m_c$, $-\f{20}{9}$ has to be replaced by
$\f{40}{9}$.

\subsection{Numerical Results for Wilson Coefficients}

        In tab.~\ref{tab:1} we give numerical results for the
coefficients $z_i$ and $y_i$ with $i=1,\ldots,6, 7V, 7A$ in the NDR
scheme, for $\Lms =200, 300, 400 \mev$, $\mu=1\gev$ and
$m_t=150\gev$. The results for the coefficients of $Q_1$--$Q_6$ are
given for completeness.  They have been analyzed in
\cite{burasetal:92d}. The numerical results for them presented here
differ slightly from the ones in \cite{burasetal:92d} because in the
present paper the effect of electroweak penguins, being of higher
order in $\alpha$, has not been included. In tab.~\ref{tab:2} we show
the $\mu$-dependence of $z_{7V}/\alpha$ for different values of $\Lms$
and for $m_t=150\gev$. The coefficient $z_{7V}$ does not depend on
$m_t$. In tab.~\ref{tab:3} we show the $\mu$-dependence of
$y_{7V}/\alpha$ for different values of $\Lms$ and for $m_t=150\gev$.
In tab.~\ref{tab:4} we show the $m_t$-dependence of $y_{7V}/\alpha$
and $y_{7A}/\alpha$ for different values of $\Lms$ and for
$\mu=1.0\gev$.  Concentrating on the coefficients of $Q_{7V}$ and
$Q_{7A}$ we observe:
\begin{itemize}
\item
The coefficient $y_{7V}$ is for $m_t=150\gev$ enhanced by 30--40\% over
the leading order result so that the suppression of this coefficient
through QCD, found in \cite{dib1:89,dib2:89,flynn:89b}, is
considerably weakened. Without QCD corrections one would find
$y_{7V}/\alpha = 0.795$. For higher values of $m_t$ the effect of
next-to-leading order corrections, albeit sizeable, is weaker:
25--35\% enhancement for $m_t=200\gev$ and smaller for higher
$m_t$. This feature will be clearly understood in the presentation in
section 5.
\item
The scheme- and $\mu$-dependences of $y_{7V}$ are very weak as shown
in particular in tab.~\ref{tab:3}. Also the dependence of $y_{7V}$ on
$\Lms$ is rather weak, at most of 3\% in the range of parameters
considered. The weak $\mu$-dependence is related to the fact that
$Q_{7V}$ does not carry any anomalous dimension and that for $\mu
<m_c$ the coefficient $y_{7V}$ is only influenced by QCD penguin
operators because $y_1=y_2=0$.
\item
The scheme- and $\mu$-dependences of $z_{7V}$ shown in
tab.~\ref{tab:2} are very strong. This implies that the calculation of
the real amplitudes such as $K_S\ra \pi^0 e^+e^-$ requires the
inclusion of the matrix elements of $Q_1$--$Q_6$ in order to cancel
the scheme- and $\mu$-dependences present in $z_{7V}$. Since the
matrix elements $\me{\pi^0e^+e^- |Q_i|K_S}$ can only be calculated
using non-perturbative methods, it is clear that $K_S\ra
\pi^0 e^+e^-$ and consequently the indirect CP violation in \kpiee are
much harder to calculate than $BR(K_L\ra \pi^0 e^+e^-)_{dir}$ .
\item
Inspecting tab.~\ref{tab:4} and fig.~\ref{fig:5} we observe in
accordance with the analyses of \cite{dib1:89,dib2:89,flynn:89b} that
at a certain value of $m_t$ the coefficient $|y_{7A}|$ becomes larger
than $|y_{7V}|$. For LO this transition takes place roughly for
$m_t=140\pm 10\gev$, where the precise value depends on $\Lms$. The
next-to-leading order corrections shift this transition point to
$m_t=180\pm 5\gev$.
\end{itemize}

\newsection{$BR(K_L \ra \pi^0 e^+ e^-)_{dir}$ Beyond Leading
Logarithms} \label{s.br}
\subsection{Analytic Formula}

        Let us introduce the numerical constant
\be
\kappa = \f{1}{V_{us}^2} \f{\tau(K_L)}{\tau(K^+)} \, BR(K^+ \ra \pi^0 e^+ \nu)
 = 4.16\, .
\ee
We find then as in \cite{dib2:89}
\be \label{final}
BR(K_L \ra \pi^0 e^+ e^-)_{dir} = \kappa (\IM\lambda_t)^2
(y_{7A}^2 + y_{7V}^2)
\ee
where
\be
\IM\lambda_t = \IM(V_{td} V^*_{ts}) \approx |V_{ub}|\ |V_{cb}| \sin\delta
\ee
in the standard parametrization of the CKM matrix.

        In deriving eq.~(\ref{final}), the matrix elements of the
operators $Q_3$--$Q_6$ have been neglected, i.~e.~we have assumed that
\be
\label{dominance}
\sum_{i=3}^6 y_i(\mu) \me{\pi^0 e^+e^- |Q_i| K_L} \ll y_{7V}(\mu) \me{\pi^0
e^+ e^- |Q_{7V}|K_L}.
\ee
This assumption is supported by the corresponding relation for the
quark-level matrix elements
\be
\sum_{i=3}^6 y_i(\mu) \me{d e^+e^- |Q_i| s} \ll y_{7V}(\mu) \me{d e^+ e^-
|Q_{7V}|s}
\ee
that can be easily verified perturbatively. The neglect of the matrix
elements of the QCD penguin oerators is compatible with the very weak
$\mu$-dependence of the retained contribution.

        Using next the method of the penguin-box expansion
\cite{buchallaetal:91} we can write
\begin{eqnarray}
\label{pbe7v}
y_{7V} & = & \f{\alpha}{2\pi \sin^2 \theta_W} \left( P_0 +
Y(x_t) - 4 \sin^2 \theta_W \; Z(x_t) + P_E \; E(x_t)\right), \\
\label{pbe7a}
y_{7A} & = & - \f{\alpha}{2\pi \sin^2 \theta_W} Y(x_t)
\end{eqnarray}
where the gauge invariant combinations of $B(x)$, $C(x)$ and $D(x)$
are given as follows:
\be
Y(x) = C(x) - B(x), \hspace{2cm} Z(x) = C(x) + \f{1}{4} D(x).
\ee
To a very good approximation one has for $120\gev \leq m_t \leq 300
\gev$ \cite{buchallaetal:91}
\be
Y(x_t) = 0.315 x_t^{0.78}, \hspace{2cm} Z(x_t) = 0.175 x_t^{0.93}.
\ee

        The constants $P_0$ and $P_E$ summarize the contributions from
scales below $\mu = M_W$. $P_E$ is ${\cal O}(10^{-2})$ and
consequently the last term in eq.~(\ref{pbe7v}) can be
neglected. $P_0$ is given for different values of $\mu$ and $\Lms$ in
tab.~\ref{tab:5}. We show there also the leading order results and
the case without QCD corrections. The analytic expressions in
eqs.~(\ref{pbe7v}) and (\ref{pbe7a}) are useful as they show not only
the explicit $m_t$-dependence, but also isolate the impact of leading
and next-to-leading QCD effects. These effects modify only the
constants $P_0$ and $P_E$. As anticipated from the results of section
\ref{s.ham}, $P_0$ is strongly enhanced relatively to the LO
result. This enhancement amounts roughly to a factor of $1.6\pm
0.1$. Part of this enhancement is however due to the fact that for
$\Lambda_{LO}=\Lms$ the QCD coupling constant in the leading order is
20--30\% larger than its next-to-leading order value. Calculating
$P_0$ in LO but with the full $\alpha_s$ of (\ref{alpha}) we have
found that the enhancement amounts then to a factor of $1.33\pm
0.06$. In any case the inclusion of NLO QCD effects and a meaningful
use of $\Lms$ show that the next-to-leading order effects weaken the
QCD suppression of $y_{7V}$. As seen in tab.~\ref{tab:5}, the
suppression of $P_0$ by QCD corrections amounts to about 15\% in the
complete next-to-leading calculation.

\figintext{
\begin{figure}[hbt]
\vspace{0.10in}
\centerline{
\epsfysize=4in
\rotate[r]{
\epsffile{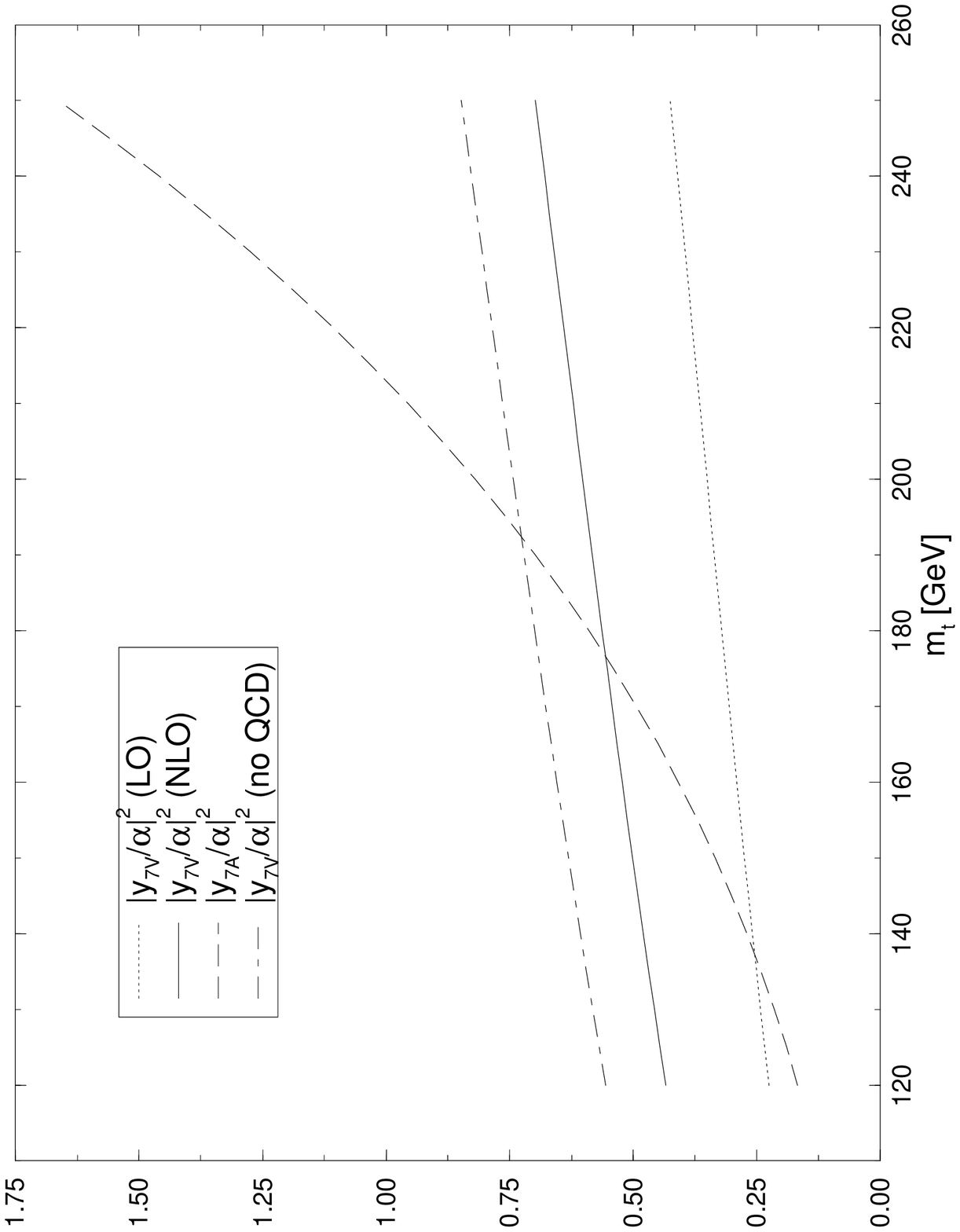}
} }
\vspace{0.08in}
\caption[]{
Wilson coefficients $|y_{7V}/\alpha|^2$ and $|y_{7A}/\alpha|^2$ as a
function of $m_t$ for ${\Lambda_{\overline{\rm MS}}}=0.3\gev$ at
scale $\mu=1.0\gev$.
\label{fig:5}}
\end{figure}
}

        We next find that
\be
Y(x_t) - 4 \sin^2 \theta_W \; Z(x_t) = \left\{ \begin{array}{ccc}
0.332 & {\rm for} & m_t=150\gev\\
0.431 & {\rm for} & m_t=200\gev\\
0.523 & {\rm for} & m_t=250\gev \end{array} \right.
\ee
whereas in the same range of $m_t$ the function $Y(x_t)$ changes from
0.840 to 1.863. Because of the additional constant term $P_0$ in
eq.~(\ref{pbe7v}), the $m_t$-dependence of $y_{7V}$ is rather weak and
consequently the dominant $m_t$-dependence of $BR(K_L\ra\pi^0 e^+
e^-)_{dir}$ for fixed $\IM\lambda_t$ originates from the coefficient
of the operator $Q_{7A}$. In fact, for $m_t > 180GeV$ this operator
gives larger contribution to the branching ratio than $Q_{7V}$. This
is shown in fig.~\ref{fig:5} where $|y_{7V}/\alpha|^2$ and
$|y_{7A}/\alpha|^2$ are plotted as functions of $m_t$ together with
the leading order result and the case without QCD
corrections. Finally, we note that for $m_t < 200 \gev$ the constant
term proportional to $P_0$ constitutes at least 60\% of $y_{7V}$.

\subsection{Uncertainties Due to the Definition of $m_t$}

        At the level of accuracy at which we work we cannot address
the question of the definition of $m_t$ used here. In order to be able
to analyze this question, one would have to calculate perturbative QCD
corrections to the functions $Y(x_t)$ and $Z(x_t)$ and include also an
additional order in the renormalization group improved perturbative
calculation of $P_0$. The latter would require evaluation of
three-loop anomalous dimension matrices, which in the near future
nobody will attempt. In any case, we expect only a small correction to
$P_0$. The uncertainty due to the choice of $\mu$ in $m_t(\mu)$ can be
substantial, as stressed in
refs.~\cite{buchallaburas:93a,buchallaburas:93b,buchallaburas:94}, and
may result in 20--30\% uncertainties in the branching ratios. It can
only be reduced if ${\cal O}(\al)$ corrections to $Y(x_t)$ and
$Z(x_t)$ are included. For $K^+\ra\pi^+\nu\bar{\nu}$,
$K_L\ra\pi^0\nu\bar{\nu}$, $B\ra\mu^+\mu^-$ and $B\ra X_s\nu\bar{\nu}$
this has been done in
refs.~\cite{buchallaburas:93a,buchallaburas:93b,buchallaburas:94}. The
inclusion of these corrections reduces the uncertainty in the
corresponding branching ratios to a few percent. Fortunately, the
result for the corrected function $Y(x_t)$ given in
refs.~\cite{buchallaburas:93a,buchallaburas:93b,buchallaburas:94} can
be directly used here. The message of
refs.~\cite{buchallaburas:93a,buchallaburas:93b,buchallaburas:94} is
the following: For $m_t = \overline{m}_t(m_t)$, the QCD corrections to
$Y(x_t)$ are below 2\%. Corresponding corrections to $Z(x)$ are not
known. Fortunately, the $m_t$-dependence of $y_{7V}$ is much
weaker and the uncertainty due to the choice of $\mu$ in $m_t(\mu)$ is
small. On the basis of these arguments and the result of
refs.~\cite{buchallaburas:93a,buchallaburas:93b,buchallaburas:94} we
believe that if $m_t = \overline{m}_t(m_t)$ is chosen, the additional
QCD corrections to $BR(K_L \ra \pi^0 e^+ e^-)_{dir}$ should be small.

\figintext{
\begin{figure}[hbt]
\vspace{0.10in}
\centerline{
\epsfysize=6.5in
\rotate[r]{
\epsffile{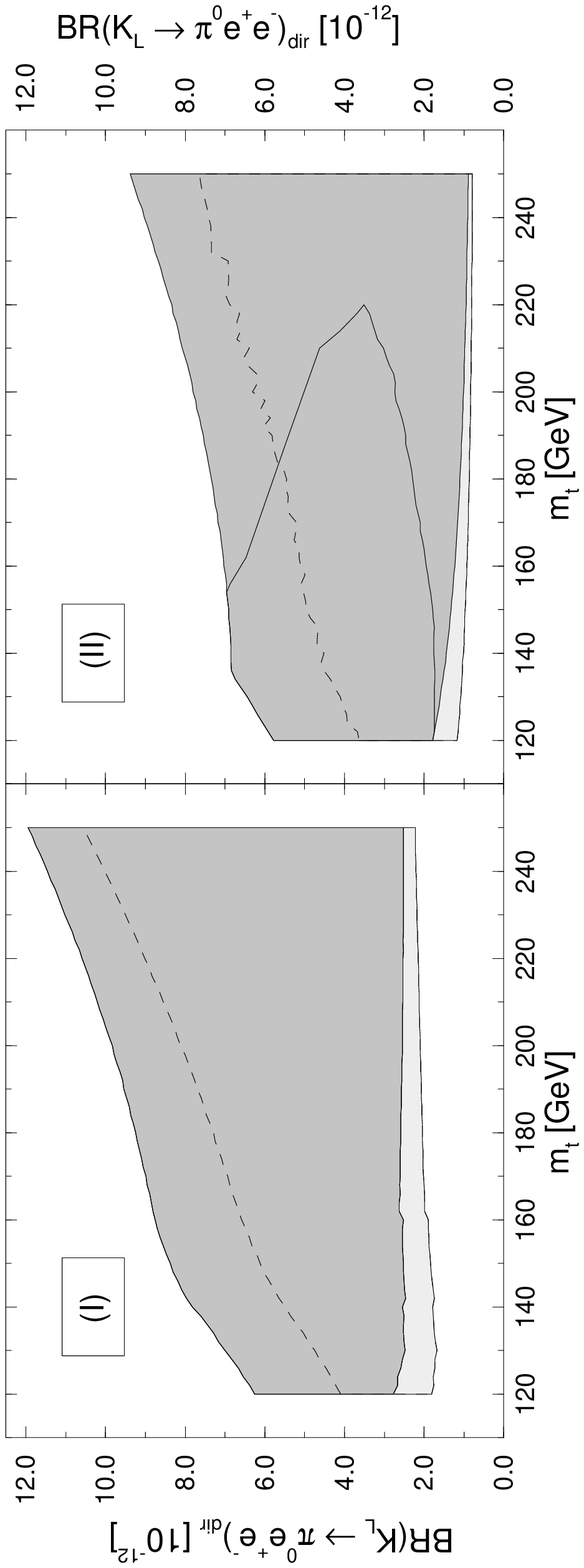}
} }
\vspace{0.08in}
\caption[]{
Allowed ranges for $BR(K_L \longrightarrow \pi^0 e^+e^-)_{\rm dir}$ as
a function of $m_t$ for ${\Lambda_{\overline{\rm MS}}}=0.3\gev$ at
scale $\mu=1.0\gev$. The labels (I) and (II) refer to the two possible
solutions for the CKM phase $\delta$. Parameters $|V_{cb}|$,
$|V_{ub}/V_{cb}|$ and $B_K$ were varied within present experimental
limits as given in (\ref{rangea}). The light and dark shaded areas
correspond to the LO and NLO results, respectively. The solid line
inside the dark shaded area describes the additional restriction from
$B^0$--$\overline{B}^0$ mixing (see the text).
\label{fig:1}}
\end{figure}
}

\subsection{Numerical Analysis}

        In the numerical calculation we need the value of $\IM
\lambda_t$ which can be extracted from the usual analysis of the
parameter $\epsilon$ that measures indirect CP violation. The formula
for $\epsilon$ can be found for instance in section 7 of
ref.~\cite{burasetal:92d} as well as in many other papers and will not
be repeated here. We only give the values of the parameters which we
use here. We take $|V_{us}| = 0.22$,
\be
\eta_1 = 1.1 \cite{herrlichnierste:93},\hspace{0.5cm} \eta_2 = 0.57
\cite{burasjaminweisz:90},\hspace{0.5cm}
\eta_3=0.36\cite{kaufmanetal:88,buchallaetal:90,dattaetal:90,flynn:90}\mbox{
(leading order)}.
\ee
and
\be
\label{rangea}
\!\mbox{Range a: }|V_{cb}| = 0.040 \pm 0.005,\ |V_{ub}/V_{cb}| = 0.08
\pm 0.02,\ B_K = 0.70\pm 0.20,
\ee

\figintext{
\begin{figure}[bt]
\vspace{0.10in}
\centerline{
\epsfysize=6.5in
\rotate[r]{
\epsffile{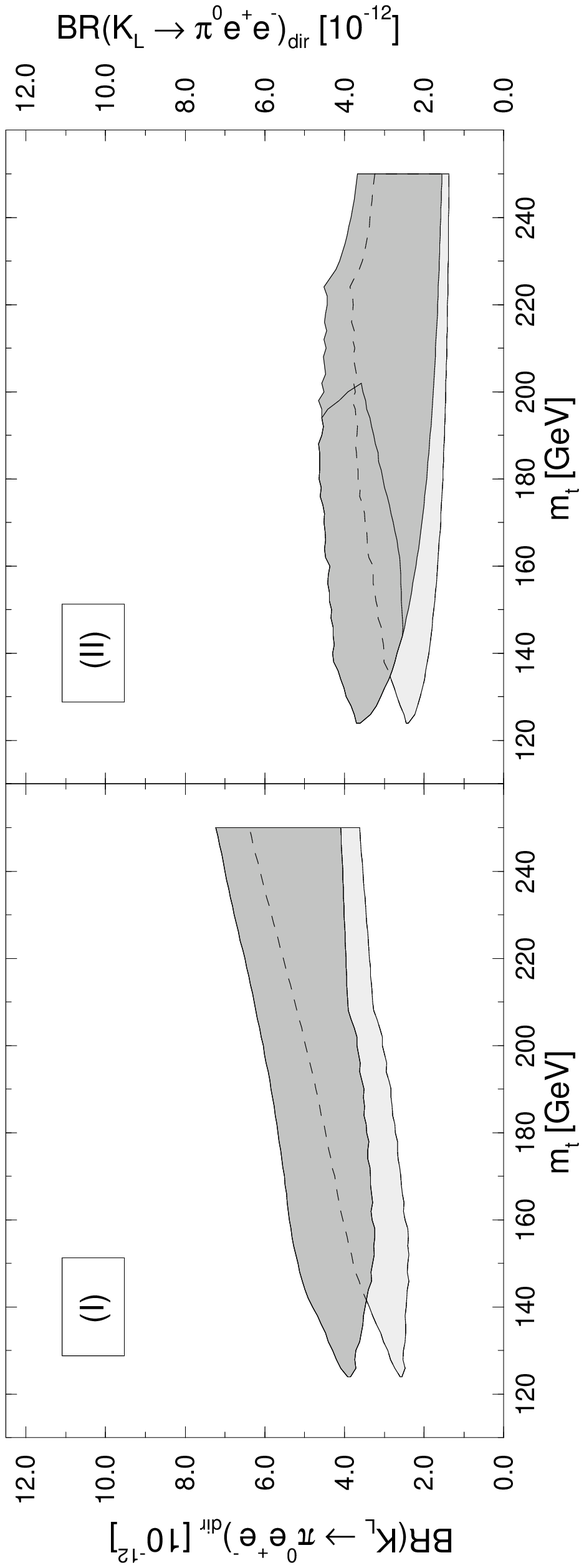}
} }
\vspace{0.08in}
\caption[]{
The same as in fig.~\ref{fig:1}, but for the parameter range given in
(\ref{rangeb}).
\label{fig:2}}
\end{figure}
}

The analysis of $\epsilon$ gives two solutions for the CKM phase
$\delta$. The values of $\IM\lambda_t$ decrease with increasing $m_t$
and this decrease is stronger for $\delta$ in the second quadrant. The
dependence of $\IM\lambda_t$ on $m_t$ compensates the corresponding
$m_t$-dependence of $y_{7V}$ and $y_{7A}$.  As a result, $BR(K_L \ra
\pi^0 e^+ e^-)_{dir}$ decreases with $m_t$ for the solution II and is
increasing with $m_t$ for the solution I substantially slower than one
would expect from eq.~(\ref{final}) would $\IM\lambda_t$ be
independent of $m_t$. In tab.~\ref{tab:6} we show the values of
$BR(K_L\ra\pi^0 e^+ e^-)_{dir}$ in units of $10^{-12}$ for $B_K=0.7$,
$|V_{cb}|=0.041$, $|V_{ub}/V_{cb}|=0.10$ as a function of $m_t$. For
$m_t \le 140\gev$, $\epsilon$ cannot be fitted for these choice of
parameters. In fig.~\ref{fig:1} we present the results when the
parameters are varied in the full range given in (\ref{rangea}). In
order to illustrate the impact of possible future improvements in the
determination of $B_K$ and the CKM parameters, in figs.~\ref{fig:2}
and \ref{fig:3} we show the results when smaller ranges of parameters
given by
\begin{eqnarray}
\label{rangeb}
\!\!\!\mbox{Range b: }|V_{cb}| = 0.040\pm 0.002,\, |V_{ub}/V_{cb}| =
0.08\pm 0.01,\, B_K = 0.75\pm 0.05 \\
\label{rangec}
\!\!\!\mbox{Range c: }|V_{cb}| = 0.043 \pm 0.002,\, |V_{ub}/V_{cb}| =
0.09 \pm 0.01,\, B_K = 0.55 \pm 0.05
\end{eqnarray}

\figintext{
\begin{figure}[hbt]
\vspace{0.10in}
\centerline{
\epsfysize=6.5in
\rotate[r]{
\epsffile{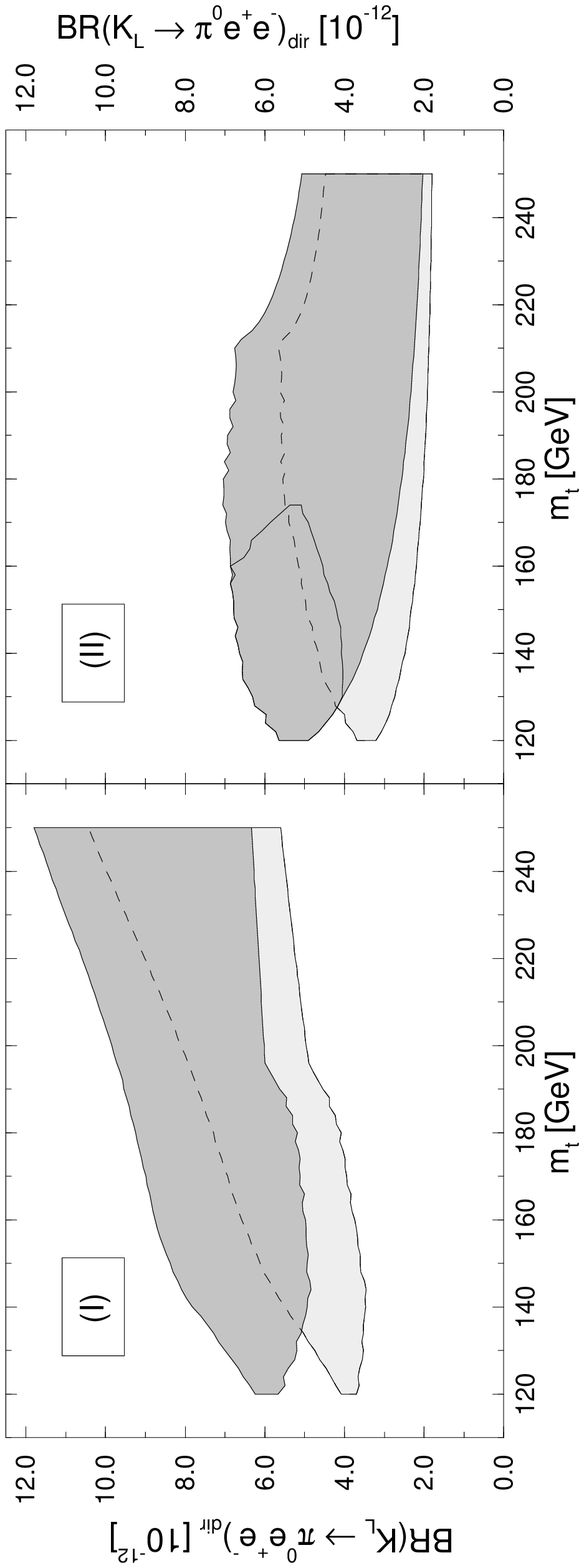}
} }
\vspace{0.08in}
\caption[]{
The same as in fig.~\ref{fig:1}, but for the parameter range given in
(\ref{rangec}).
\label{fig:3}}
\end{figure}
}

are used. $B_K$ in range b is in the ball park of most recent lattice
$(B_K=0.825\pm 0.027\pm 0.023)$ \cite{kilcupetal:93} and $1/N$
$(B_K=0.7\pm0.1)$ \cite{bardeenetal:88} results. In range c it is
closer to the values obtained in the hadron duality approach
$(B_K=0.4\pm 0.1)$ \cite{pradesetal:91}. In the latter case we had to
take larger values for $|V_{cb}|$ in order to obtain solutions in the
$\epsilon$-analysis.

\figintext{\begin{figure}[hbt]
\vspace{0.10in}
\centerline{
\epsfysize=4in
\rotate[r]{
\epsffile{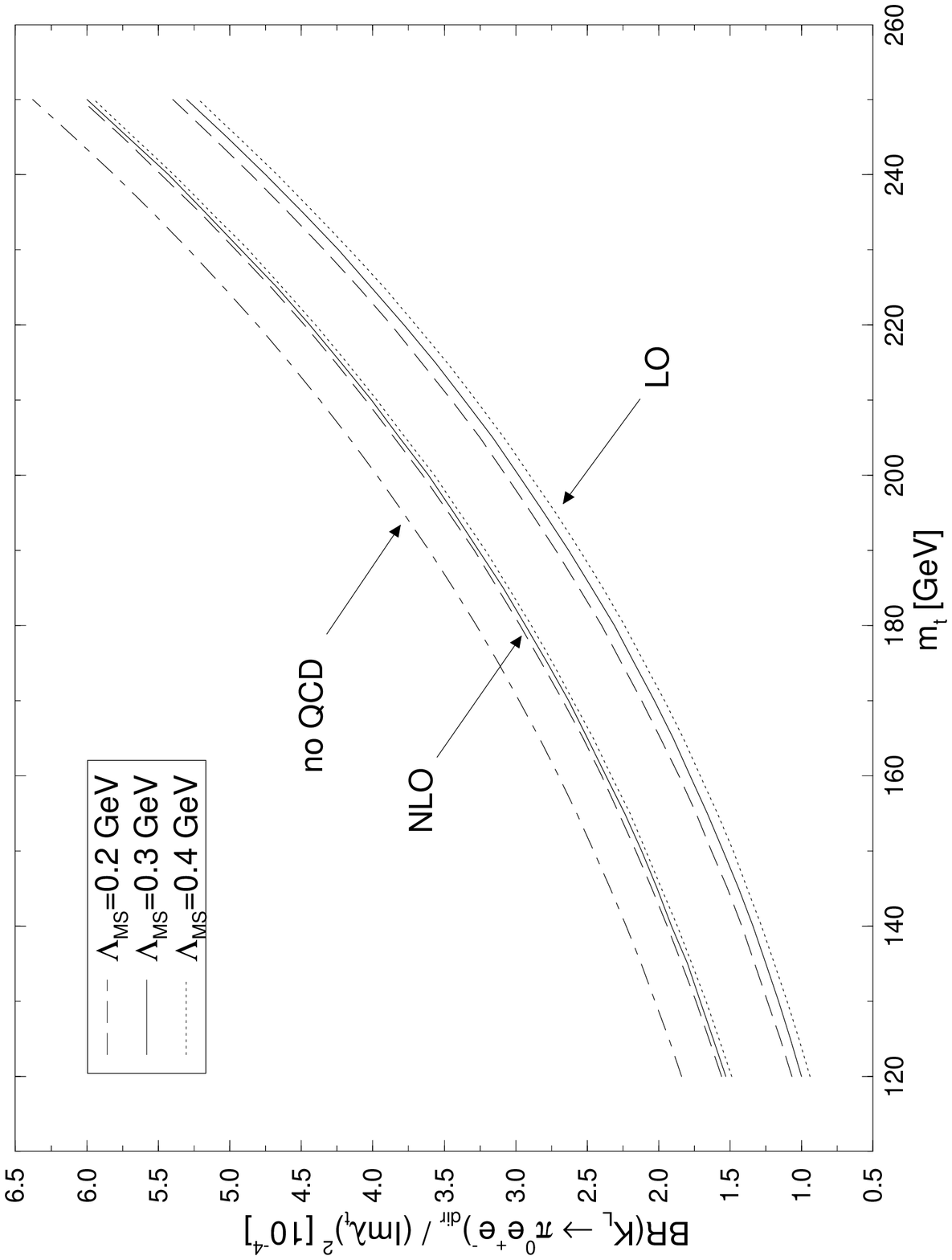}
} }
\vspace{0.08in}
\caption[]{
$BR(K_L \longrightarrow \pi^0 e^+e^-)_{\rm dir}/(\IM\lambda_t)^2$
as a function of $m_t$ for various values of $\Lms$ at scale
$\mu =1.0\gev $.
\label{fig:4}}
\end{figure}
}

	Finally in fig.~\ref{fig:4} we show the ratio
$BR(K_L\ra\pi^0 e^+ e^-)/(\IM\lambda_t)^2$ as a function of $m_t$.
All these results are self explanatory and we only make a few
comments:
\begin{itemize}
\item
The NLO corrections enhance the direct CP violating contributions with
respect to its LO estimate.
\item
Due to large uncertainties in $\IM\lambda_t$ which enters the branching
ratio quadratically, the enhancement found here cannot be appreciated
yet. However as seen in figs.~\ref{fig:2} and \ref{fig:3} the more
accurate data for $\IM\lambda_t$ in the future will allow to feel this
enhancement clearly.
\item
For central values of the parameters in range a and $m_t=170\pm
30\gev$ we find
\be
BR(K_L\ra\pi^0 e^+ e^-)_{dir} = \left\{\begin{array}{ll}
(4.9\pm 0.5)\cdot 10^{-12}, & \mbox{Solution I} \\
(3.7\pm 0.5)\cdot 10^{-12}, & \mbox{Solution II,} \end{array}\right.
\ee
but as seen in fig.~\ref{fig:1} values as high as $10^{-11}$ are not
excluded when the solution I is chosen.
\end{itemize}

        The analysis of $B^0$--$\bar{B}^0$ mixing can constrain
further the predicted values of $BR($\kpiee$)_{dir}$. The usual box
diagrams with top quark exchanges give for the mixing parameter
\be
x_d = 3.89\cdot 10^3 \left[\f{\tau_B}{1.5 \, {\rm ps}}\right]\left[
\f{B_B F_B^2}{(200\mev)^2}\right] |V_{td}|^2 S(x_t),
\ee
where
\be
S(x) = \f{x}{4}\left( 1 + \f{9}{1-x} - \f{6}{(1-x)^2}\right) +
\f{3x^3}{2(x-1)^3} \ln x.
\ee
We use
\be
\sqrt{B_B}\; F_B = 200\pm 30\mev
\ee
which is in the ball park of various lattice and QCD sum rule estimates
\cite{kronfeldmackenzie:93}. Setting $\tau_B = 1.5 \, {\rm ps}$ we
require
\be
x_d = 0.72 \pm 0.08\,\cite{cassel:93}
\ee
in accordance with the most recent average of CLEO and ARGUS data. The
impact of $B^0$--$\overline{B}^0$ mixing with $\sqrt{B_B}F_B$ as
given above is shown in figs.~\ref{fig:1}--\ref{fig:3} for the NLO
results as a solid line within the dark shaded area. With this
additional requirement the region to the right of the solid line is
excluded. The constraint is invisible for solution~I. For solution~II
we observe that for higher values of $m_t$ the constraint coming from
$B^0$--$\overline{B}^0$ mixing is rather effective. These features are
in accordance with general expectations that for high values of $m_t$
and $F_B$ the solution I is favoured and that with more precise high
values of $F_B$ the solution II could even be excluded. This would
also constrain the values for $BR($\kpiee$)_{dir}$ for solution~I.

\newsection{Comparison with the Other Two Contributions} \label{s.compar}
        Now we want to compare the results obtained in the previous
section with the estimates made for the indirect CP-violating
contribution and the CP-conserving one. The most recent summary of the
present status has been given by Pich \cite{pich:93}. We have nothing
to add to this issue here and only state his final conclusions.

        The indirect CP violating contribution is given by the
$K_S\ra\pi^0 e^+ e^-$ amplitude times the CP parameter $\epsilon$.
Once $BR(K_S\ra\pi^0 e^+ e^-)$ has been accurately measured, it will
be possible to calculate this contribution precisely. Using chiral
perturbation theory it is however possible to get an estimate by
relating $K_S\ra\pi^0 e^+ e^-$ to the $K^+\ra\pi^+ e^+ e^-$
transition. Present data implies
\be
BR(K_L\ra\pi^0 e^+ e^-)_{indir} \le 1.6\cdot 10^{-12},
\ee
i.~e. a branching ratio more than a factor of 2 below the direct CP
violating contribution for the CKM phase $\delta$ in the second
quadrant. More importantly, if the solution I is chosen, the direct
CP violating contribution is by a factor of 5 or more larger than the
indirect one. This should be contrasted with $K\ra\pi\pi$ decays. It
should be stressed, however, that in reality the CP indirect amplitude
may interfere with the vector part of the CP direct amplitude
\footnote{The axial part of the CP indirect amplitude can be safely
neglected}. This interference could have both signs. For a very heavy
top quark for which the first quadrant is favoured and the
axial-vector contribution to the direct CP violation dominates, the
presence of the indirect CP violation will not have a large impact on
our analysis given above. Yet for $m_t={\cal O}(150\gev)$ and $\delta$
in the second quadrant, the indirect CP violation cannot be neglected
and a more careful estimate of the full CP violating branching ratio
requires a better analysis which includes the interference in question.

        The estimate of the CP conserving contribution is more
difficult. We refer the reader to ref.~\cite{pich:93} where the
literature on this subject can be found. The most recent analysis in
the framework of chiral perturbation theory gives \cite{cohenetal:93}
\be
BR(K_L\ra\pi^0 e^+ e^-)_{cons} \approx (0.3-1.8)\cdot 10^{-12},
\ee
i.~e.~well below the {\it direct} CP violating contribution. An
improved estimate of this component of \kpiee is certainly desirable.

\newsection{Summary and Outlook} \label{s.sum}
        We have calculated the direct CP violating contribution to
\kpiee beyond the leading logarithmic approximation. This required the
evaluation of two-loop mixing between the four-quark operators
$Q_1$--$Q_6$ and the semileptonic operator $Q_{7V}$. Our main findings
can be summarized as follows:
\begin{itemize}
\item
The next-to-leading QCD corrections enhance the branching ratio
$BR(K_L\ra\pi^0 e^+ e^-)_{dir}$ so that its suppression found in the
leading order analysis is considerably weakened.
\item
The net QCD suppression of this contribution for $m_t>150\gev$ is at
most 20\%.
\item
The very weak dependences of the resulting $BR(K_L\ra\pi^0 e^+
e^-)_{dir}$ on $\Lms$ and $\mu$ indicate that this branching ratio
can be reliably calculated in the Standard Model, provided $m_t$ and
the CKM parameters have been better determined.
\item With the help of the two-loop calculation of
refs.~\cite{buchallaburas:93a,buchallaburas:93b,buchallaburas:94} we
have also some control over the renormalization scale of $m_t$.
Consequently when $m_t=\overline{m}_t(m_t)$ is chosen we expect on the
basis of
refs.~\cite{buchallaburas:93a,buchallaburas:93b,buchallaburas:94} that
still higher order QCD corrections are very small.
\item
Provided the phase $\delta$ is in the first quadrant and $m_t=170\pm
30\gev$ we expect
\be
\label{brranges}
BR(K_L\ra\pi^0 e^+ e^-)_{dir} = \left\{\begin{array}{ll} (6.0\pm
4.0)\cdot 10^{-12}, & \mbox{Range a} \\
(4.5\pm 1.5)\cdot 10^{-12}, & \mbox{Range b} \\
(7.5\pm 2.5)\cdot 10^{-12}, & \mbox{Range c} \end{array}\right.
\ee for the three ranges of parameters considered. As seen from
(\ref{brranges}) the present prediction, range a, will be improved
when the CKM parameters and $B_K$ have been better determined and the
top quark has been discovered. For $\delta$ in the second quadrant the
branching ratio is typically by a factor of two smaller.
\item
These results confirm the expectations made previously by other
authors that the decay \kpiee is dominated by direct CP violation
(see \cite{pich:93} and references therein). The present experimental
bounds
\be
BR(K_L\ra\pi^0 e^+ e^-) \leq\left\{ \begin{array}{ll}
4.3 \cdot 10^{-9} & \cite{harrisetal:93} \\
5.5 \cdot 10^{-9} & \cite{ohletal:90} \end{array} \right.
\ee
are still by three orders of magnitude away from the theoretical
expectations in the Standard Model. Yet the prospects of getting the
required sensitivity of order $10^{-11}$--$10^{-12}$ in five years are
encouraging
\cite{winsteinwolfenstein:93,ritchiewojcicki:93,littenbergvalencia:93}.
\end{itemize}

\bigskip
\begin{center}
{\large\bf Acknowledgment}
\end{center}
\noindent
M.~Misiak would like to thank A.~Pich for helpful discussions.

\newpage
\begin{appendix}
\noindent {\Large\bf Appendices}\\
\newsection{Calculation in the HV Scheme}
\label{HV}
        The Wilson coefficients in the HV scheme can be found using
\be
\vec{C}_{HV}(\mu)=\left[\hat{\bf 1} -
\f{\alpha_s(\mu)}{4\pi}\Delta \hat{r}^T\right]\vec{C}_{NDR}(\mu),
\ee
where $\Delta \hat{r}$ has been defined in (\ref{me}). The $6\times 6$
submatrix of $\Delta \hat{r}$ relating HV and NDR schemes is given in
eqs.~(3.9)--(3.13) of ref.~\cite{burasetal:92a}. Performing an analogous
calculation in the presence of $Q_{7V}'$ one finds the additional
non-vanishing elements of $\Delta \hat{r}$:
\be
\label{delta_r}
(\Delta \hat{r})_{17}  = -(\Delta \hat{r})_{47} = \f{8}{9} N,\quad
(\Delta \hat{r})_{27}=-(\Delta \hat{r})_{37}=\f{8}{9},\quad (\Delta
\hat{r})_{77}=2C_F.
\ee
The last result can be traced back to the non-vanishing anomalous
dimension $\gamma_J$ of the weak current at the two-loop level in the
HV scheme. Consequently in this scheme the operators $Q_{7V}$ and
$Q_{7A}$ carry anomalous dimensions at the two-loop level. Similarly
the anomalous dimensions of the operators $Q_1$--$Q_6$ receive
additional large contributions which are related to $\gamma_J\ne0$ in
the HV scheme. This feature introduces an unnecessarily large scheme
dependence at the lower end of the renormalization group evolution. In
the spirit of \cite{burasetal:92d} we perform therefore an additional
finite renormalization which removes these additional contributions to
the anomalous dimensions of $Q_1$--$Q_6$ and makes the anomalous
dimensions of $Q_{7V}$ and $Q_{7A}$ in the HV scheme to be zero. This
transformation is equivalent to the following changes in $\Delta
\hat{r}$ of (\ref{delta_r}):
\begin{eqnarray}
\label{6x6}
(\Delta \hat{r})_{6\times 6} & \ra & (\Delta \hat{r})_{6\times 6} - 4
C_F \hat{\bf 1}_{6\times 6}, \\
\label{77element}
(\Delta \hat{r})_{77} & \ra & (\Delta \hat{r})_{77}-2C_F = 0.
\end{eqnarray}
The shift in the element (7,7) differs from the shift in (\ref{6x6}).
Therefore, according to eq.~(\ref{rendep}) this finite renormalization
modifies also the mixing of the operators $Q_1$--$Q_6$ with
$Q_{7V}'$. Using (\ref{rendep}), the matrix $\hat{\gamma}^{(1)}$
calculated in the NDR scheme and $\Delta \hat{r}$ discussed above
we find $\hat{\gamma}^{(1)}$ in the HV scheme modified by the finite
transformations (\ref{6x6}) and (\ref{77element}). Here we give only the
seventh column of $\hat{\gamma}_{HV}^{(1)}$:
\begin{eqnarray}
\left(\gamma_{17}^{(1)}\right)_{HV} & = & \f{8}{3}\left(1-N^2\right),
\hspace{1.5cm}
\left(\gamma_{27}^{(1)}\right)_{HV} = \f{520}{81}\left(\f{1}{N}-
N\right), \\
\left(\gamma_{37}^{(1)}\right)_{HV} & = & \f{8}{3} \left(
u-\f{d}{2}\right) \left( 1-N^2\right) + \f{256}{81} \left(
N-\f{1}{N}\right), \\
\left(\gamma_{47}^{(1)}\right)_{HV} & = & \left( \f{520}{81}u-
\f{128}{81}d \right)\left(\f{1}{N}-N\right) + \f{8}{3}\left( N^2-1
\right), \\
\left(\gamma_{57}^{(1)}\right)_{HV} & = & \f{8}{3} \left(
u-\f{d}{2}\right)\left( 1-N^2\right), \\
\left(\gamma_{67}^{(1)}\right)_{HV} & = &
\left(\f{376}{81}u-\f{56}{81}d \right)\left(
\f{1}{N}-N\right), \\
\left(\gamma_{77}^{(1)}\right)_{HV} & = & -2 \beta_1.
\end{eqnarray}
We observe that $\gamma_{17}^{(1)}$, $\gamma_{57}^{(1)}$ and
$\gamma_{77}^{(1)}$ are the same in both schemes.

        Finally we give the initial conditions in the HV scheme:
\be
C_1(M_W) = \f{7}{2} \f{\alpha_s(M_W)}{4\pi}, \hspace{2cm} C_2(M_W) = 1-
\f{7}{6} \f{\alpha_s(M_W)}{4\pi}.
\ee
The other coefficients are given in eqs.~(\ref{c3-c6})--(\ref{BCDE})
with the constant pieces in (\ref{tildefunc}) removed.

\newsection{Evanescent Differences}
\label{mikolaj}
        As described in section \ref{sb.gamma1}, the two-loop
anomalous dimension matrix entries $\gamma^{(1)}_{17}$ --
$\gamma^{(1)}_{67}$ in eqs.~(\ref{17,27})--(\ref{67}) have been
extracted from the calculations presented in
refs.~\cite{burasetal:92a,burasetal:92b,burasetal:92c}.  In this
appendix, we would like to show that all the $\gamma^{(1)}_{17}$ --
$\gamma^{(1)}_{67}$ found here agree with the mixings found
independently by one of us in ref.~\cite{misiak:93}, after the
differences in the conventions used are taken into account.

        The operators responsible for the $b \ra s e^+ e^-$ decay
considered in ref.~\cite{misiak:93} differ from the ones in
eq.~(\ref{oper}) of the present paper only by changes in flavours and
by trivial normalization factors.  The changes in flavours are
irrelevant for the values of the mixings, as long as the numbers of
active up- and down-quarks are kept as free variables, which was the
case in both papers.  In the present paper, we have followed the
normalization conventions of
refs.~\cite{burasweisz:90,burasetal:92a,burasetal:92b,burasetal:92c}.
In consequence, the operators $Q_1$--$Q_6$ in eq.~(\ref{oper}) here
are four times larger than $O_1$ -- $O_6$ in eq.~(3) of
ref.~\cite{misiak:93}. The operator $Q'_{7V}$ here is two times larger
than $O_9$ defined in eqs.~(3) and (6) of
ref.~\cite{misiak:93}. Comparing in addition eqs.~(\ref{expgam}) here
and (20) in ref.~\cite{misiak:93} one verifies that
\be
\begin{array}{cccc} \label{compar}
\gamma^{(0)}_{i7} &=& 4 B^1_{i3}, & i=1,\ldots,6, \\
\gamma^{(1)}_{i7} &=& 4 \left( B^2_{i3} + \Delta B^2_{i3} \right),
&i=1,\ldots,6
\end{array}
\ee
where the quantities on the l.h.s are the ones from
eqs.~(\ref{17,27:LO})--(\ref{57,67:LO}) and (\ref{17,27})--(\ref{67})
of the present paper, while $B^1_{i3}$ and $B^2_{i3}$ are taken from
eq.~(21) of ref.~\cite{misiak:93} with the following substitutions
made:
\be
 \overline{Q} = uQ_u+dQ_d = \f{2}{3} \left( u - \f{d}{2} \right),
\;\;\;\;\;\;\;\;\;\;\;\;\;\;\;\;
 f=u+d  \;\;\;\;\;\;\;\;\;\;{\rm and}\;\;\;\;\;\; \xi = 0.
\ee
Finally, $\Delta B^2_{i3}$ is the shift required in order to take into
account the difference between the renormalization schemes used. In
the following, we will check that its value is exactly what is required for
eq.~(\ref{compar}) to hold.

        In the present calculation we have used the results of
refs.~\cite{burasetal:92a,burasetal:92b,burasetal:92c} obtained within the
$\overline{MS}$ scheme
and NDR regularization. This was also the scheme used in
ref.~\cite{misiak:93}.  However, in order to fully specify a scheme, one also
has to fix certain convention-dependent constants in the definitions
of the evanescent operators, whenever such operators enter the
game.  Generalizing eq.~(15) of ref.~\cite{misiak:93}, one can write the
evanescent operators relevant in our calculation in the following
form:
\be \label{defev}
O^{ev}_{i+10} \equiv \f{1}{6} O_i \left( \gamma_{\mu} \ra
\gamma_{[\mu} \gamma_{\nu} \gamma_{\rho]} \right) \pm [1 + b_i (4-D)]
O_i, \;\;\;\;\;\; i=1,\ldots,6
\ee
with ``$-$'' for $i=1,\ldots,4$ and ``$+$'' for $i=5,6$. D denotes the
dimensionality of spacetime. The numbers $b_i$ are arbitrary. Each
choice of them corresponds to a different renormalization scheme.

        In ref.~\cite{misiak:93} all the $b_i$'s were set to zero. On the
other hand, in the papers
\cite{burasweisz:90,burasetal:92a,burasetal:92b,burasetal:92c} they differed
from
zero, and they were fixed by requiring that certain projections of the
evanescent operators vanish. For $i=1,\ldots,4$, the relevant identity
was (see eq.~(4.3) of ref.~\cite{burasetal:92b})\footnote{ We neglect
irrelevant terms of order ${\cal O} ((4-D)^2)$ in eqs.~(\ref{defev}),
(\ref{14}) and (\ref{56}) here.}
\be \label{14}
\begin{array}{ccc}
0= E_{\alpha \beta, \gamma \delta}
\left( \gamma_{\tau} (1 + \gamma_5) \right)_{\beta \gamma}
\left( \gamma^{\tau} (1 + \gamma_5) \right)_{\delta \alpha}
&=& 2 Tr \{ \f{1}{6}
\gamma_{[\mu} \gamma_{\nu} \gamma_{\rho]} \gamma_{\tau}
\gamma^{[\mu} \gamma^{\nu} \gamma^{\rho]} \gamma^{\tau} (1 + \gamma_5)-\\
\ & \ & \ \\
- [1 + b_i (4-D)]
\gamma_{\mu} \gamma_{\tau} \gamma^{\mu} \gamma^{\tau} (1 + \gamma_5)\}
&=& 64 (4-D) (b_i - \f{1}{6} ).
\end{array}
\ee
Eq. (4.8) of ref.~\cite{burasetal:92b} gives the corresponding requirement for
$i=5,6$:
\be \label{56}
\begin{array}{ccc}
0= E_{\alpha \beta, \gamma \delta} (1 - \gamma_5)_{\beta \gamma}
(1 + \gamma_5)_{\delta \alpha}
&=& 2 Tr \{ \f{1}{6}
\gamma_{[\mu} \gamma_{\nu} \gamma_{\rho]}
\gamma^{[\mu} \gamma^{\nu} \gamma^{\rho]} (1 + \gamma_5)+ \\
\ & \ & \ \\
+ [1 + b_i (4-D)]
\gamma_{\mu} \gamma^{\mu} (1 + \gamma_5)\}
&=& 32 (4-D) (b_i + \f{5}{6} ).
\end{array}
\ee
Consequently, the conventions of
refs.~\cite{burasweisz:90,burasetal:92a,burasetal:92b,burasetal:92c}
correspond to taking
\be \label{bs}
b_1 = b_2 = b_3 = b_4 = \f{1}{6}, \hspace{3cm} b_5 = b_6 = -\f{5}{6}.
\ee

	In the following, we will use the notation of ref.~\cite{misiak:93} to
find the explicit values of $\Delta B^2_{i3}$ for $i=1,\ldots,6$.

	According to eq.~(16) of ref \cite{misiak:93}, the
contribution of the evanescent operators to the two-loop anomalous
dimension matrix for the ``usual'' operators depends on the
divergences they renormalize (matrix $a^{11}_{ik}$) and on their
finite one-loop matrix elements (matrix $a^{01}_{kj}$). The matrix
$a^{11}_{ik}$ is insensitive to the ${\cal O}(4-D)$ parts of the
evanescent operators, i.e. it is insensitive to shifts in the numbers
$b_i$. However, the finite one-loop matrix elements of the evanescent
operators depend on these numbers. If one changes $b_i$ from zero to
any nonzero value, then the corresponding change in the one-loop
matrix element of $O^{ev}_{i+10}$ is proportional to the product of
$b_i$ and the divergent part of the one-loop matrix element of the
``normal'' operator that is multiplied by $b_i$ in the definition of
$O^{ev}_{i+10}$. Explicitly:
\be
\Delta a^{01}_{(i+10)j} = a^{11}_{ij} (\pm b_i).
\ee
Consequently, according to eqs.~(8), (16), (20) and (21) of
ref.~\cite{misiak:93}, one gets
\be
\Delta B^2_{i3} = 16 \pi^2 \sum_{k=1}^{6} a^{11}_{i(k+10)} B^1_{k3} (\pm
b_k), \;\;\;\;\;\;\; i=1,\ldots,6
\ee
with ``$-$'' for $i=1,\ldots,4$ and ``$+$'' for $i=5,6$. Taking the
numbers $b_i$ from eq.~(B.6) above, and the matrices $a^{11}$ and
$B^1$ from eqs.~(17) and (21) of ref.~\cite{misiak:93}, respectively, one
obtains the final result for $\Delta B^2_{i3}$
\be
\Delta B^2_{i3} = \left[\f{2}{3} N c_F Q_u,\;\; \f{4}{3} c_F Q_u,\;\;
\f{8}{3} c_F Q_d, \;\; \f{4}{3}c_F \overline{Q} + \f{4}{3} N c_F Q_d,\;\;
0, \;\; \f{20}{3} c_F \overline{Q} \right]_i
\ee
which is exactly what is needed for eq.~(\ref{compar}) to hold.

        We have thus shown that the mixings of the four-quark
operators into the quark-lepton operator found in
ref.~\cite{misiak:93} are in agreement with the ones found here on the
basis of the results of
refs.~\cite{burasetal:92a,burasetal:92b,burasetal:92c}, as long as the
differences in the conventions are properly taken into account. This
suggests that the disagreement between refs.~\cite{misiak:93} and
\cite{cellaetal:91} (see app.~B of ref.~\cite{misiak:93}) should be
resolved in favour of ref.~\cite{misiak:93}. However, we have to
point out that the next-to-leading matching conditions and mixings of
the four-quark operators from ref.~\cite{burasetal:92a} have not been
used in a fully consistent manner in ref.~\cite{misiak:93}. As long as
one wants to define the evanescent operators as in
ref.~\cite{misiak:93} (all the numbers $b_i$ vanishing), one has to
transform accordingly both the next-to-leading terms in the matching
conditions for the the four-quark operators $Q_1$ and $Q_2$, and the
whole $6 \times 6$ next-to-leading anomalous dimension matrix of the
four-quark operators that has been calculated in
ref.~\cite{burasetal:92a}. The author of ref.~\cite{misiak:93} was not
aware of the fact that the conventions used in
ref.~\cite{burasetal:92a} for the evanescent operators were different
from his own. In consequence, the numerical results of
ref.~\cite{misiak:93} for the $b\ra s\,e^+\,e^-$ decay are somewhat
larger than they should be. The change is however phenomenologically
irrelevant since it never exceeds 3\% .

\newsection{Calculation of $\gamma_{37}^{(1)}$ and $\gamma_{47}^{(1)}$ in
the NDR Scheme}
\label{bases}
The procedure described below constitutes a synthesis of subsection
5.1 of refs.~\cite{burasetal:92b,burasetal:92c}. The calculation is
more complicated than in the pure QCD case of
ref.~\cite{burasetal:92b} because here we only deal with pure ${\cal
O}(\al^2)$ anomalous dimension matrices after a proper rescaling of
$Q_{7V}$.

In order to solve the problem of closed fermion loops involving $\gamma_5$
in the NDR scheme, we have to consider four different bases of
operators.
All four bases contain the penguin operators $Q_3$ -- $Q'_{7V}$ of
eq.~(\ref{oper}) but differ in the current--current operators $Q_1$ and
$Q_2$. The latter are given as follows,
\begin{equation}
\hbox{\bf Basis A:\ }
Q_1^{(u)} = \left(\bar s_\alpha u_\beta  \right)_{V-A}
                  \left(\bar u_\beta  d_\alpha \right)_{V-A} \,,
\quad
Q_2^{(u)} = \left(\bar s u \right)_{V-A}
                  \left(\bar u d \right)_{V-A} \,.
\end{equation}
\begin{equation}
\hbox{\bf Basis B:\ }
Q_1^{(d)} = \left(\bar s_\alpha d_\beta  \right)_{V-A}
                  \left(\bar d_\beta  d_\alpha \right)_{V-A} \,,
\quad
Q_2^{(d)} = \left(\bar s d \right)_{V-A}
                  \left(\bar d d \right)_{V-A} \,.
\end{equation}
\begin{equation}
\hbox{\bf Basis C:\ }
\widetilde{Q}_1^{(u)} = \left(\bar s d \right)_{V-A}
                              \left(\bar u u \right)_{V-A} \,,
\quad
\widetilde{Q}_2^{(u)} = \left(\bar s_\alpha d_\beta  \right)_{V-A}
                             \left(\bar u_\beta  u_\alpha \right)_{V-A}
\,.
\end{equation}
\begin{equation}
\hbox{\bf Basis D:\ }
\widetilde{Q}_1^{(d)} = \left(\bar s d \right)_{V-A}
                              \left(\bar d d \right)_{V-A} \,,
\quad
\widetilde{Q}_2^{(d)} = \left(\bar s_\alpha d_\beta  \right)_{V-A}
                             \left(\bar d_\beta  d_\alpha \right)_{V-A}
\,.
\end{equation}

The basis A is the standard basis of eq.~(\ref{oper}) and the basis B is
an auxiliary basis needed for the solution of the problem. The bases
C and D are simply Fierz conjugates of $Q_1$ and $Q_2$ in A and B,
respectively. Evidently,
\begin{equation}
\label{equivalence}
Q_1^{(d)} \; = \; \widetilde{Q}_2^{(d)} \, ,
\qquad
Q_2^{(d)} \; = \; \widetilde{Q}_1^{(d)} \, .
\end{equation}

Let us next denote by $[Q_i]_1$ and $[Q_i]_2$ the result of the type-1
(closed fermion loop) and type-2 insertions (no closed fermion loop) of
an operator $Q_i$, respectively. Then the results for penguin
contributions to the row entries of $\gamma^{(1)}$ for $Q_3$ and $Q_4$
can be written as follows
\begin{eqnarray}
\label{Q3}
{[Q_3]_{\rm p}} &=& u\,[\widetilde{Q}_1^{(u)}]_1 +
              d\,[\widetilde{Q}_1^{(d)}]_1 +
2\,[\widetilde{Q}_1^{(d)}]_2 \,, \\
\label{Q4}
{[Q_4]_{\rm p}} &=& u\,[\widetilde{Q}_2^{(u)}]_1 +
              d\,[\widetilde{Q}_2^{(d)}]_1 +
2\,[\widetilde{Q}_2^{(d)}]_2 \,.
\end{eqnarray}
For the calculation of the entries in the first six columns it is not
necessary to introduce the bases B and D because in pure QCD no
distinction is made between $Q_{1,2}^{(u)}$ and $Q_{1,2}^{(d)}$. This
was in fact the case considered in ref.~\cite{burasetal:92b}. However,
here the seventh column involves couplings to the photon and as in
ref.~\cite{burasetal:92c} the bases B and D have to be introduced.

There are six independent entries in eq.~(\ref{Q3}) and eq.~(\ref{Q4})
which have to be found. In the following four steps the quoted results
for these entries are given for the seventh column only.

\leftline{\bf Step 1:}\noindent
$[\widetilde{Q}_1^{(d)}]_2$ and $[\widetilde{Q}_2^{(d)}]_2$ can be
calculated without any problems by evaluating the diagrams of
fig.~\ref{diag:5}.  The contributions to the seventh column are given by
\begin{eqnarray}
[\widetilde{Q}_1^{(d)}]_2 &=& -\,\frac{464}{81}\,C_F \,,\\
{[\widetilde{Q}_2^{(d)}]_2} &=&  -\,\frac{1}{2}\,[Q_1^{(u)}]_2 \; = \;
                                 \frac{8}{3}\,N\,C_F \,.
\end{eqnarray}

\leftline{\bf Step 2:}\noindent
$\widetilde{Q}_1^{(u)}$ and $\widetilde{Q}_2^{(u)}$ receive
contributions only from type-1 insertions. This allows to find
$[\widetilde{Q}_1^{(u)}]_1$ and $[\widetilde{Q}_2^{(u)}]_1$ by
comparing the two--loop anomalous dimension matrices calculated in the
bases C and A using the relation

\begin{equation}
\big(\hat{\gamma}^{(1)}\big)_p^{(C)} \; = \;
\big(\hat{\gamma}^{(1)}\big)_p^{(A)} +
\left[ \Delta\hat{r},\hat{\gamma}^{(0)} \right] + 2\,\beta_0\, \Delta\hat{r}
\,,
\end{equation}
where
\begin{equation}
\Delta\hat{r} = \left[ \hat{r} \right]_{C} - \left[ \hat{r} \right]_{A}
\end{equation}
with $\hat{r}$ defined in eq.~(\ref{me}).

	Because the insertion in the current-current diagrams are
identical for these two bases, only penguin diagram contributions to
$\hat{\gamma}^{(1)}$ and $\Delta\hat{r}$ enter this relation. On the
other hand $\hat{\gamma}^{(0)}$ is the full one-loop matrix discussed
in section~\ref{sb.gamma0}. A simple calculation of finite terms in
one-loop penguin diagrams gives
\begin{equation}
\Delta\hat{r} =
-\,\frac{1}{3}\,
\left( \begin{array}{c} 0 \\ 1 \\ 0 \\ \vdots \\ 0 \end{array} \right) \, P+
\frac{8}{9}\,
\left( \begin{array}{c} N \\ 1 \\ 0 \\ \vdots \\ 0 \end{array} \right) \,
\widetilde{P} \,,
\end{equation}
where
\begin{eqnarray}
P & = & (0, 0, -1/N, 1, -1/N, 1, 0), \nonumber \\
\widetilde{P} &=& (0, 0, 0, 0, 0, 0, 1).
\end{eqnarray}
In summary, we find
\begin{equation}
\left[ \widetilde{Q}_1^{(u)} \right]_1 \; = \; \frac{8}{3} \, (1-N^2) \,,
\qquad
\left[ \widetilde{Q}_2^{(u)} \right]_1 \; = \;
   \frac{280}{81} \, (\frac{1}{N}-N) \,.
\end{equation}

\leftline{\bf Step 3:}\noindent
The contribution of $[\widetilde{Q}_1^{(d)}]_1$ to the seventh column
can be related to the corresponding contribution of
$[\widetilde{Q}_1^{(u)}]_1$ found in step 2.  Inspecting the diagrams
of fig.~\ref{diag:5} we obtain
\begin{equation}
[\widetilde{Q}_1^{(d)}]_1 \; = \;
-\,\frac{1}{2}\,[\widetilde{Q}_1^{(u)}]_1 \; = \; \frac{4}{3} \, (N^2 -1) \,.
\end{equation}

\leftline{\bf Step 4:}\noindent
The calculation of $[\widetilde{Q}_2^{(d)}]_1$ proceeds as follows. Since
$\widetilde{Q}_2^{(d)}$ receives contributions from both type-1 and type-2
insertions we can write
\begin{equation}
[\widetilde{Q}_2^{(d)}]_1 \; = \; [\widetilde{Q}_2^{(d)}]_{\rm p} -
[\widetilde{Q}_2^{(d)}]_2 \,,
\end{equation}
with the last entry calculated in step 1.

In order to find $[\widetilde{Q}_2^{(d)}]_{\rm p}$ we compare the
two--loop
anomalous dimension matrices calculated in the bases D and B using
the relation
\begin{equation}\
\label{transformationBD}
\big(\hat{\gamma}^{(1)}\big)_p^{(D)} \; = \;
\big(\hat{\gamma}^{(1)}\big)_p^{(B)} +
\left[ (\Delta\hat{r})^{(d)}, (\hat{\gamma}^{(0)})^{(d)} \right] +
2\,\beta_0\, \Delta\hat{r}^{(d)} \,,
\end{equation}
where the index $d$ indicates that now the auxiliary bases D and B are
considered. The matrix $(\hat{\gamma}^{(0)})^{(d)}$ differs from
$(\hat{\gamma}^{(0)})$ only in the first row and the entry $(2,7)$. The
first row in $(\hat{\gamma}^{(0)})^{(d)}$ equals the second row and
\begin{equation}
(\gamma^{(0)}_{17})^{(d)} =
(\gamma^{(0)}_{27})^{(d)} = \frac{8}{9}\, (N+1) \,.
\end{equation}
$\hat{\gamma}^{(1)}$ and $(\Delta\hat{r})^{(d)}$ involve again only the
penguin contributions. From one-loop penguin diagrams we find
\begin{equation}
(\Delta\hat{r})^{(d)} =
\frac{1}{3}\,
\left( \begin{array}{c} 1 \\ -1 \\ 0 \\ \vdots \\ 0 \end{array}
\right) \, P - \frac{4}{9}\,(N-1)\,
\left( \begin{array}{c} 1 \\ -1 \\ 0 \\ \vdots \\ 0 \end{array} \right) \,
\widetilde{P} \,.
\end{equation}
In this way the last two terms in eq.~(\ref{transformationBD}) can be
evaluated.  Since both types of insertions of $\widetilde{Q}_1^{(d)}$
have been calculated in steps 1 and 3, the element $\left[
\widetilde{Q}_2^{(d)}
\right]_1$ can finally be extracted from eq.~(\ref{transformationBD}) when in
addition the relation (\ref{equivalence}) is used. The result is
\begin{equation}
\left[ \widetilde{Q}_2^{(d)} \right]_1 \; = \;
\frac{64}{81} \, (\frac{1}{N}-N) \,.
\end{equation}

Putting all this together and using eqs.~(\ref{Q3}) and (\ref{Q4}) we
find the results for $\gamma_{37}^{(1)}$ and $\gamma_{47}^{(1)}$ for the
NDR scheme quoted in eqs.~(\ref{37}) and (\ref{47}), respectively.

\newsection{Collection of Numerical Input Parameters}
\label{parameters}
\bigskip

\leftline{\large\bf Quark Masses}
\begin{displaymath}
m_t = 120\;-\;250\gev \qquad \qquad m_b = 4.8\gev \qquad \qquad
m_c = 1.4\gev
\end{displaymath}

\smallskip

\leftline{\large\bf QCD and Electroweak Parameters}
\begin{displaymath}
\begin{array}{rclrcl}
\Lms & = & 200\mev : & \al(M_Z) & = & 0.109 \\
\Lms & = & 300\mev : & \al(M_Z) & = & 0.116 \\
\Lms & = & 400\mev : & \al(M_Z) & = & 0.122 \\
\alpha & = & 1/128 & G_F & = & 1.16639\cdot 10^{-5}\gev^{-2} \\
\sin^2\theta_W & = & 0.23 & M_W & = & 80.0\gev
\end{array}
\end{displaymath}

\smallskip

\leftline{\large\bf CKM Elements and $B_K$ Ranges}
\begin{displaymath}
\begin{array}{llclcl}
& \left| V_{us}\right| = 0.22 &&& \\
{\rm Range~(a):} &
\left| V_{cb} \right| = 0.040 \pm 0.005 &\ &
\left| V_{ub}/V_{cb} \right| = 0.08 \pm 0.02 &\ &
B_{\rm K} = 0.70 \pm 0.20 \\
{\rm Range~(b):} &
\left| V_{cb} \right| = 0.040 \pm 0.002 &\ &
\left| V_{ub}/V_{cb} \right| = 0.08 \pm 0.01 &\ &
B_{\rm K} = 0.75 \pm 0.05 \\
{\rm Range~(c):} &
\left| V_{cb} \right| = 0.043 \pm 0.002 &\ &
\left| V_{ub}/V_{cb} \right| = 0.09 \pm 0.01 &\ &
B_{\rm K} = 0.55 \pm 0.05 \\
\end{array}
\end{displaymath}

\smallskip

\leftline{\large\bf $K$ Decays and $K^0-\bar{K}^0$ Mixing}
\begin{displaymath}
\begin{array}{lrclrcl}
BR(K^+ \rightarrow \pi^0 e^+\nu)  =  0.0482 \quad & m_{\rm K} & = &
498\mev & \eta_1 & = & 1.1 \\
\tau_{K_L} = 5.17   \cdot 10^{-8}\,s     & F_{\rm K} & = & 161\mev &
\eta_2 & = & 0.57 \\
\tau_{K^+} = 1.237 \cdot 10^{-8}\,s\quad & \Delta M_{\rm K} & = & 3.5
\cdot 10^{-15}\gev \quad & \eta_3 & = & 0.36 \\
\phantom{Xi}\epsilon = 2.258 \cdot 10^{-3} & & \\
\end{array}
\end{displaymath}

\smallskip
\leftline{\large\bf $B^0-\bar{B}^0$ Mixing}
\begin{displaymath}
\begin{array}{rclrclrcl}
x_d & = & 0.72 \pm 0.08\qquad & \sqrt{B_B}\; F_B & = & 200\pm 30\mev\qquad &
 \eta_{2B} & = & 0.55
\end{array}
\end{displaymath}

\clearpage

\newsection{Tables}

\begin{table}[h]
\caption{Wilson coefficients at $\mu=1\gev$ for $\mt=150\gev$.
In our approach $y_1 = y_2 \equiv 0$ holds.\newline}
\label{tab:1}
\begin{center}
\begin{tabular}{|c||c|c|c||c|c|c||c|c|c|}
\hline
& \multicolumn{3}{c||}{$\Lms=0.2\gev$} &
  \multicolumn{3}{c||}{$\Lms=0.3\gev$} &
  \multicolumn{3}{c| }{$\Lms=0.4\gev$} \\
\hline
Scheme & LO & NDR & HV & LO &
NDR & HV & LO & NDR & HV \\
\hline
\hline
$z_1$ & -0.583 & -0.394 & -0.474 & -0.710 &
-0.483 & -0.602 & -0.848 & -0.584 & -0.769 \\
$z_2$ & 1.310 & 1.196 & 1.248 & 1.398 &
1.254 & 1.337 & 1.500 & 1.325 & 1.462 \\
\hline
$z_3$ & 0.004 & 0.008 & 0.004 & 0.005 &
0.013 & 0.008 & 0.007 & 0.021 & 0.015 \\
$z_4$ & -0.010 & -0.023 & -0.011 & -0.013 &
-0.034 & -0.018 & -0.018 & -0.054 & -0.029 \\
$z_5$ & 0.003 & 0.006 & 0.003 & 0.004 &
0.007 & 0.004 & 0.006 & 0.009 & 0.006 \\
$z_6$ & -0.010 & -0.023 & -0.010 & -0.015 &
-0.035 & -0.016 & -0.020 & -0.055 & -0.026 \\
\hline
$z_{7V}/\aem$ & -0.015 & -0.008 & 0.009 & -0.026 &
-0.037 & 0.000 & -0.037 & -0.071 & -0.009 \\
\hline
\multicolumn{9}{c}{} \\
\hline
$y_3$ & 0.027 & 0.022 & 0.024 & 0.033 &
0.028 & 0.032 & 0.041 & 0.035 & 0.042 \\
$y_4$ & -0.047 & -0.043 & -0.045 & -0.055 &
-0.051 & -0.054 & -0.063 & -0.059 & -0.064 \\
$y_5$ & 0.011 & 0.004 & 0.012 & 0.012 &
0.001 & 0.015 & 0.013 & -0.008 & 0.018 \\
$y_6$ & -0.078 & -0.071 & -0.065 & -0.101 &
-0.098 & -0.086 & -0.129 & -0.141 & -0.117 \\
\hline
$y_{7V}/\aem$ & 0.554 & 0.718 & 0.711 & 0.524 &
0.708 & 0.700 & 0.498 & 0.698 & 0.688 \\
$y_{7A}/\aem$ & -0.579 & -0.579 & -0.579 & -0.579 &
-0.579 & -0.579 & -0.579 & -0.579 & -0.579 \\
\hline
\end{tabular}
\end{center}
\end{table}

\begin{table}[thb]
\caption{Wilson coefficient $z_{7V}/\aem$ for $\mt=150\gev$ for
various values of $\mu$.
\label{tab:2}}
\begin{center}
\begin{tabular}{|c||c|c|c||c|c|c||c|c|c|}
\hline
& \multicolumn{3}{c||}{$\Lms=0.2\gev$} &
  \multicolumn{3}{c||}{$\Lms=0.3\gev$} &
  \multicolumn{3}{c| }{$\Lms=0.4\gev$} \\
\hline
$\mu [\gev]$ & LO & NDR & HV & LO &
NDR & HV & LO & NDR & HV \\
\hline
\hline
0.8 & -0.030 & -0.020 & 0.010 & -0.052 &
-0.067 & -0.007 & -0.075 & -0.127 & -0.021 \\
0.9 & -0.021 & -0.013 & 0.010 & -0.037 &
-0.049 & -0.002 & -0.053 & -0.094 & -0.015 \\
1.0 & -0.015 & -0.008 & 0.009 & -0.026 &
-0.037 & 0.000 & -0.037 & -0.071 & -0.009 \\
1.1 & -0.009 & -0.005 & 0.007 & -0.017 &
-0.028 & 0.002 & -0.025 & -0.055 & -0.005 \\
1.2 & -0.005 & -0.004 & 0.005 & -0.010 &
-0.023 & 0.002 & -0.015 & -0.044 & -0.002 \\
\hline
\end{tabular}
\end{center}
\end{table}

\begin{table}[thb]
\caption{Wilson coefficient $y_{7V}/\aem$ for $\mt=150\gev$ for
various values of $\mu$.
\label{tab:3}}
\begin{center}
\begin{tabular}{|c||c|c|c||c|c|c||c|c|c|}
\hline
& \multicolumn{3}{c||}{$\Lms=0.2\gev$} &
  \multicolumn{3}{c||}{$\Lms=0.3\gev$} &
  \multicolumn{3}{c| }{$\Lms=0.4\gev$} \\
\hline
$\mu [\gev]$ & LO & NDR & HV & LO &
NDR & HV & LO & NDR & HV \\
\hline
\hline
0.8 & 0.558 & 0.722 & 0.715 & 0.529 &
0.712 & 0.704 & 0.503 & 0.700 & 0.691 \\
0.9 & 0.556 & 0.720 & 0.713 & 0.526 &
0.710 & 0.702 & 0.500 & 0.700 & 0.690 \\
1.0 & 0.554 & 0.718 & 0.711 & 0.524 &
0.708 & 0.700 & 0.498 & 0.698 & 0.688 \\
1.1 & 0.552 & 0.717 & 0.709 & 0.523 &
0.706 & 0.698 & 0.496 & 0.696 & 0.686 \\
1.2 & 0.551 & 0.715 & 0.708 & 0.521 &
0.705 & 0.696 & 0.494 & 0.694 & 0.684 \\
\hline
\end{tabular}
\end{center}
\end{table}

\begin{table}[thb]
\caption{Wilson coefficients $y_{7V}/\aem$ (at $\mu=1\gev$) and
$y_{7A}/\aem$ for various values of $\mt$.
\label{tab:4}}
\begin{center}
\begin{tabular}{|c||c||c|c||c|c||c|c||c|}
\hline
& \multicolumn{7}{c||}{$y_{7V}/\aem$} & $y_{7A}/\aem$ \\
\hline
& & \multicolumn{2}{c||}{$\Lms=0.2\gev$}  &
    \multicolumn{2}{c||}{$\Lms=0.3\gev$}  &
    \multicolumn{2}{c|| }{$\Lms=0.4\gev$} & \\
\hline
$\mt [\gev]$ & {\rm No QCD} & LO & NDR & LO & NDR & LO & NDR & \\
\hline
\hline
130 & 0.763 & 0.522 & 0.686 &  0.493 & 0.676 & 0.466 & 0.666 & -0.464 \\
150 & 0.795 & 0.554 & 0.718 &  0.524 & 0.708 & 0.498 & 0.698 & -0.579 \\
170 & 0.823 & 0.583 & 0.747 &  0.553 & 0.737 & 0.527 & 0.727 & -0.703 \\
190 & 0.849 & 0.609 & 0.774 &  0.580 & 0.763 & 0.553 & 0.753 & -0.836 \\
210 & 0.874 & 0.634 & 0.799 &  0.605 & 0.788 & 0.578 & 0.778 & -0.978 \\
230 & 0.898 & 0.658 & 0.823 &  0.629 & 0.812 & 0.602 & 0.802 & -1.129 \\
250 & 0.921 & 0.681 & 0.846 &  0.652 & 0.836 & 0.626 & 0.825 & -1.290 \\
\hline
\end{tabular}
\end{center}
\end{table}

\begin{table}[h]
\caption{PBE coefficient $P_0$ of $y_{7V}$ for various values of
$\Lms$ and $\mu$. In the absence of QCD $P_0=8/9\;\sin^2\theta_W\ln
(M_W/m_c) = 0.827$ holds universally.
\label{tab:5}}
\begin{center}
\begin{tabular}{|cc||c|c|c|}
\hline
\multicolumn{2}{|c||}{}     &
\multicolumn{3}{c|}{$P_0$} \\
\hline
$\Lms [\gev]$ & $\mu [\gev]$ & {\rm LO} & {\rm NDR} & {\rm HV} \\
\hline
\   & 0.8 & 0.487 &  0.725 &  0.714 \\
0.2 & 1.0 & 0.482 &  0.719 &  0.709 \\
\   & 1.2 & 0.477 &  0.715 &  0.704 \\
\hline
\   & 0.8 & 0.446 &  0.711 &  0.699 \\
0.3 & 1.0 & 0.440 &  0.705 &  0.693 \\
\   & 1.2 & 0.434 &  0.700 &  0.687 \\
\hline
\   & 0.8 & 0.408 &  0.694 &  0.681 \\
0.4 & 1.0 & 0.401 &  0.690 &  0.676 \\
\   & 1.2 & 0.396 &  0.684 &  0.670 \\
\hline
\end{tabular}
\end{center}
\end{table}

\begin{table}[h]
\caption{$BR(K_{\rm L} \rightarrow \pi^0\,e^+\,e^-)_{\rm dir}
\cdot 10^{12}$ at $\mu=1.0$ for $V_{cb}=0.041$, $V_{ub}/V_{cb}=0.10$,
$B_K=0.7$ and various values of $\Lms$ and $\mt$. The labels (I)
and (II) denote the two different solutions for the CKM phase
$\delta$.
\label{tab:6}}
\begin{center}
\begin{tabular}{|cc||c|c||c|c||c|c|}
\hline

& & \multicolumn{2}{c||}{\rm LO}     &
    \multicolumn{2}{c||}{\rm NDR}    &
    \multicolumn{2}{c| }{\rm HV}     \\
\hline
$\Lms [\gev]$ & $\mt [\gev]$ & (I) & (II) & (I) & (II) & (I) & (II) \\
\hline
\  & 140 & 4.01 & 2.95 & 5.48 & 4.02 & 5.40 & 3.97 \\
\  & 160 & 4.70 & 2.44 & 6.08 & 3.15 & 6.01 & 3.11 \\
0.2 & 180 & 5.18 & 2.20 & 6.41 & 2.72 & 6.35 & 2.70 \\
\  & 200 & 5.61 & 2.07 & 6.69 & 2.46 & 6.63 & 2.44 \\
\  & 220 & 6.02 & 1.98 & 6.96 & 2.29 & 6.91 & 2.28 \\
\  & 240 & 6.41 & 1.93 & 7.24 & 2.18 & 7.20 & 2.17 \\
\hline
\  & 140 & 3.79 & 2.79 & 5.38 & 3.95 & 5.29 & 3.89 \\
\  & 160 & 4.49 & 2.33 & 5.98 & 3.10 & 5.91 & 3.06 \\
0.3 & 180 & 5.00 & 2.12 & 6.32 & 2.69 & 6.25 & 2.66 \\
\  & 200 & 5.45 & 2.01 & 6.61 & 2.44 & 6.55 & 2.41 \\
\  & 220 & 5.87 & 1.93 & 6.90 & 2.27 & 6.84 & 2.26 \\
\  & 240 & 6.28 & 1.89 & 7.18 & 2.16 & 7.13 & 2.15 \\
\hline
\  & 140 & 3.61 & 2.65 & 5.27 & 3.87 & 5.18 & 3.81 \\
\  & 160 & 4.32 & 2.24 & 5.89 & 3.05 & 5.80 & 3.01 \\
0.4 & 180 & 4.84 & 2.06 & 6.24 & 2.65 & 6.16 & 2.62 \\
\  & 200 & 5.30 & 1.95 & 6.54 & 2.41 & 6.47 & 2.38 \\
\  & 220 & 5.74 & 1.89 & 6.83 & 2.25 & 6.77 & 2.23 \\
\  & 240 & 6.17 & 1.86 & 7.12 & 2.15 & 7.07 & 2.13 \\
\hline
\end{tabular}
\end{center}
\end{table}

\clearpage

\figatend{\newsection{Feynman Diagrams}
\begin{figure}[h]}{
\vspace{0.15in}
\centerline{
\epsfysize=3.5in
\epsffile{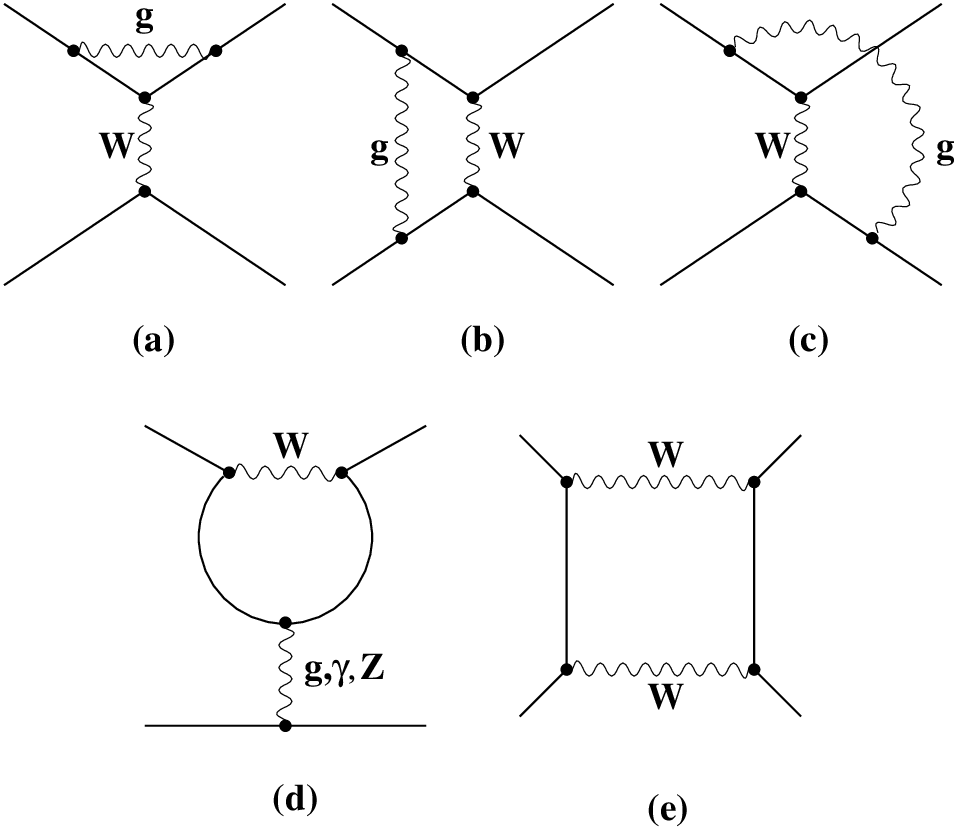}
}
\vspace{0.15in}
}{
\caption[]{
One-loop current-current and penguin diagrams in the full theory.
\label{diag:2}}
\end{figure}
}

\figatend{\begin{figure}[h]}{
\vspace{0.15in}
\centerline{
\epsfysize=2in
\epsffile{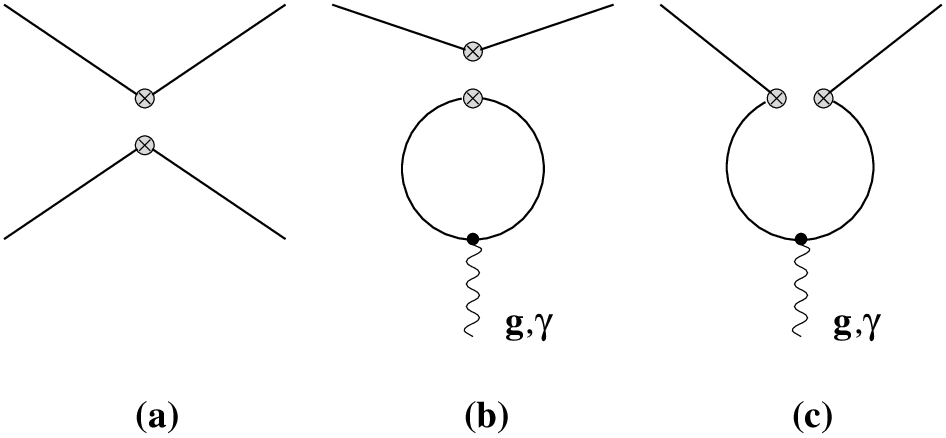}
}
\vspace{0.15in}
}{
\caption[]{
The three basic ways of inserting a given operator into a four-point
function: (a) current-current-, (b) type-1 penguin-, (c) type-2
penguin-insertion. The 4-vertices ``$\otimes\ \otimes$'' denote
standard operator insertions.
\label{diag:1}}
\end{figure}
}

\figatend{\begin{figure}[h]}{
\vspace{0.15in}
\centerline{
\epsfysize=3.5in
\epsffile{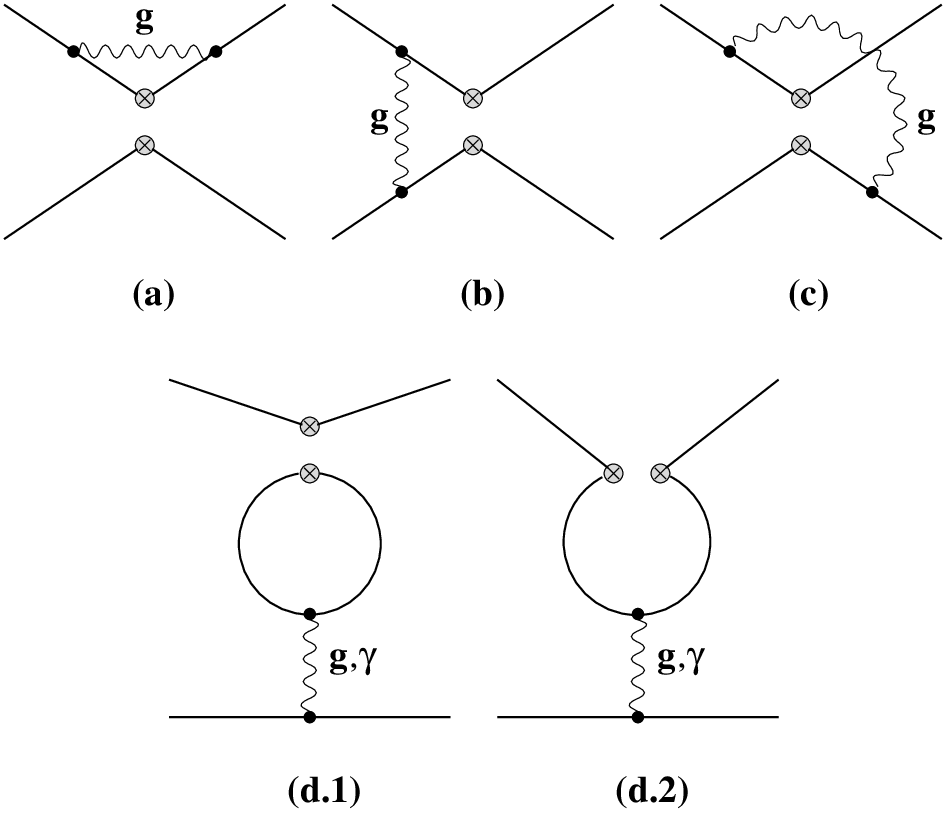}
}
\vspace{0.15in}
}{
\caption[]{
One-loop current-current and penguin diagrams contributing to
$\gamma^{(0)}$ in the effective theory. The unlabeled wavy lines denote
either a gluon or a photon. The meaning of vertices is the same as in
fig~\ref{diag:1}. Possible left-right or up-down reflected diagrams are
not shown.
\label{diag:3}}
\end{figure}
}

\figatend{\begin{figure}[h]}{
\vspace{0.15in}
\centerline{
\epsfysize=4in
\epsffile{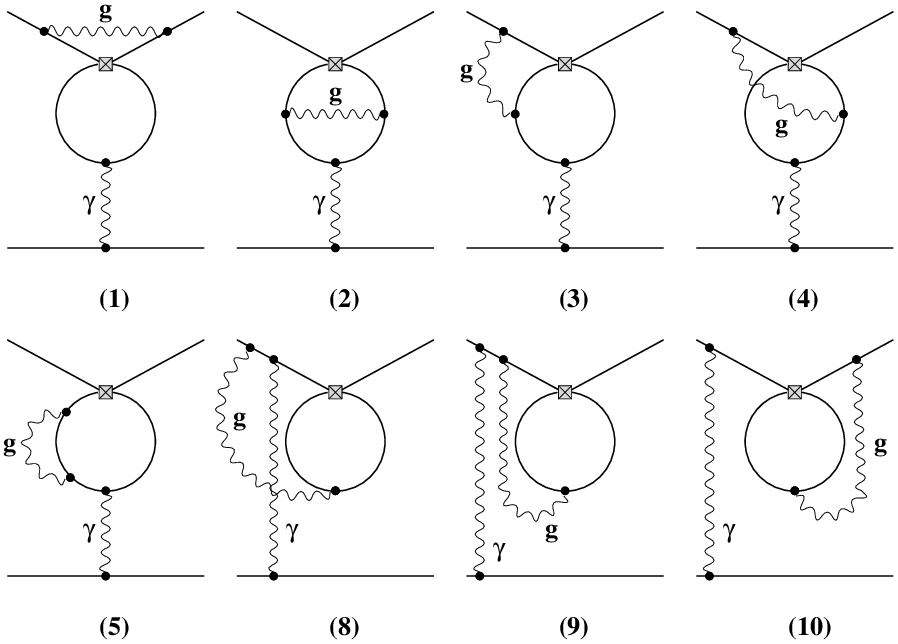}
}
\vspace{0.15in}
}{
\caption[]{
Two--loop penguin diagrams contributing to $\gamma^{(1)}_{i7}$,
$i=1,\ldots,6$. Square vertices stand for type-1 and type-2 penguin
insertions as of figs.~\ref{diag:1}(b) and (c), respectively. Possible
left-right reflected diagrams are not shown. The numbering of the
diagrams corresponds to the notation in ref.~\cite{burasetal:92b}.
\label{diag:5}}
\end{figure}
\clearpage
}

\figatend{
\goodbreak
\section{Figures}
\begin{figure}[h]}{
\vspace{0.15in}
\centerline{
\epsfysize=4in
\rotate[r]{
\epsffile{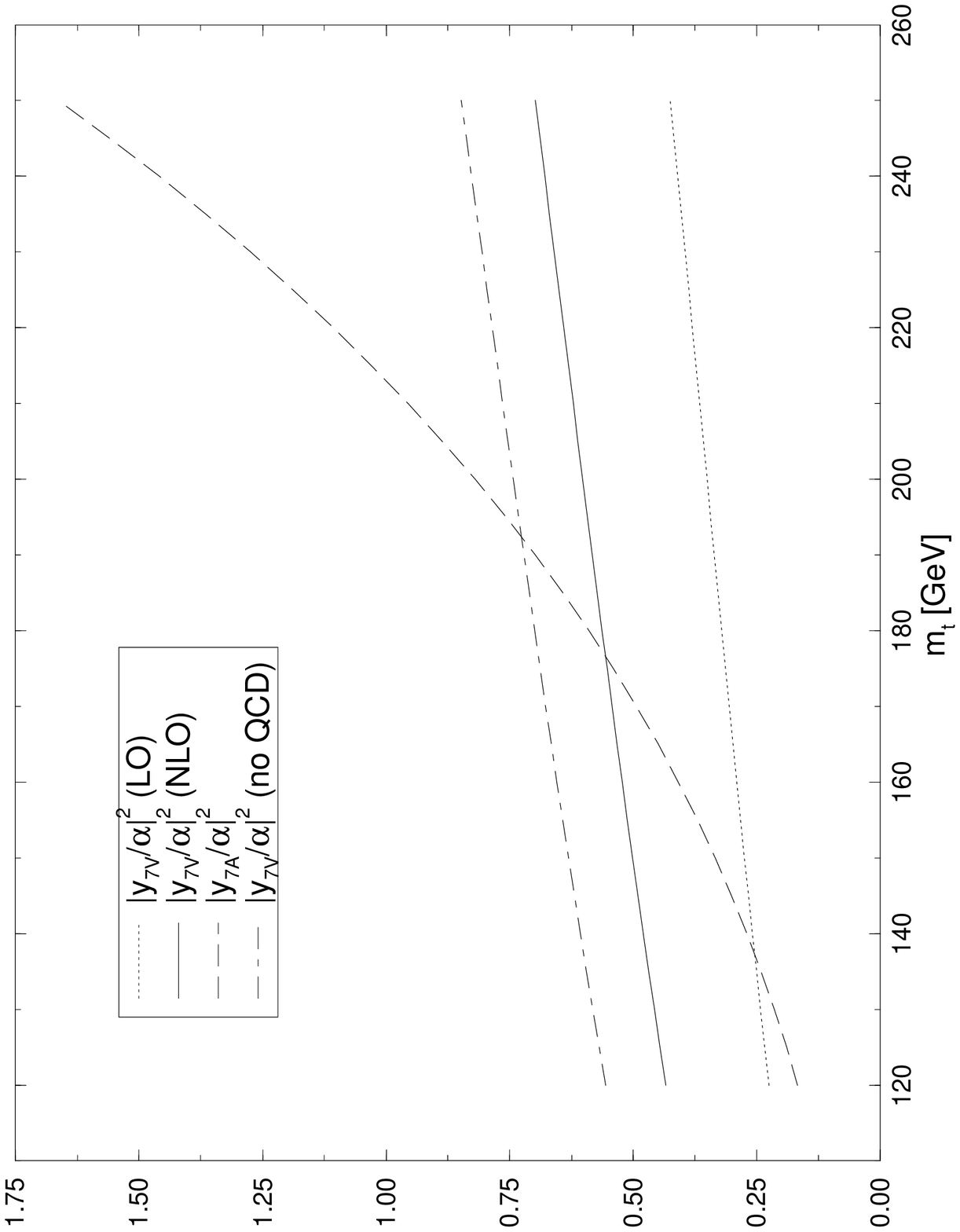}
} }
\vspace{0.15in}
}{
\caption[]{
Wilson coefficients $|y_{7V}/\alpha|^2$ and $|y_{7A}/\alpha|^2$ as a
function of $m_t$ for ${\Lambda_{\overline{\rm MS}}}=0.3\gev$ at
scale $\mu=1.0\gev1$.
\label{fig:5}}
\end{figure}
}

\figatend{\begin{figure}[h]}{
\vspace{0.15in}
\centerline{
\epsfysize=6.5in
\rotate[r]{
\epsffile{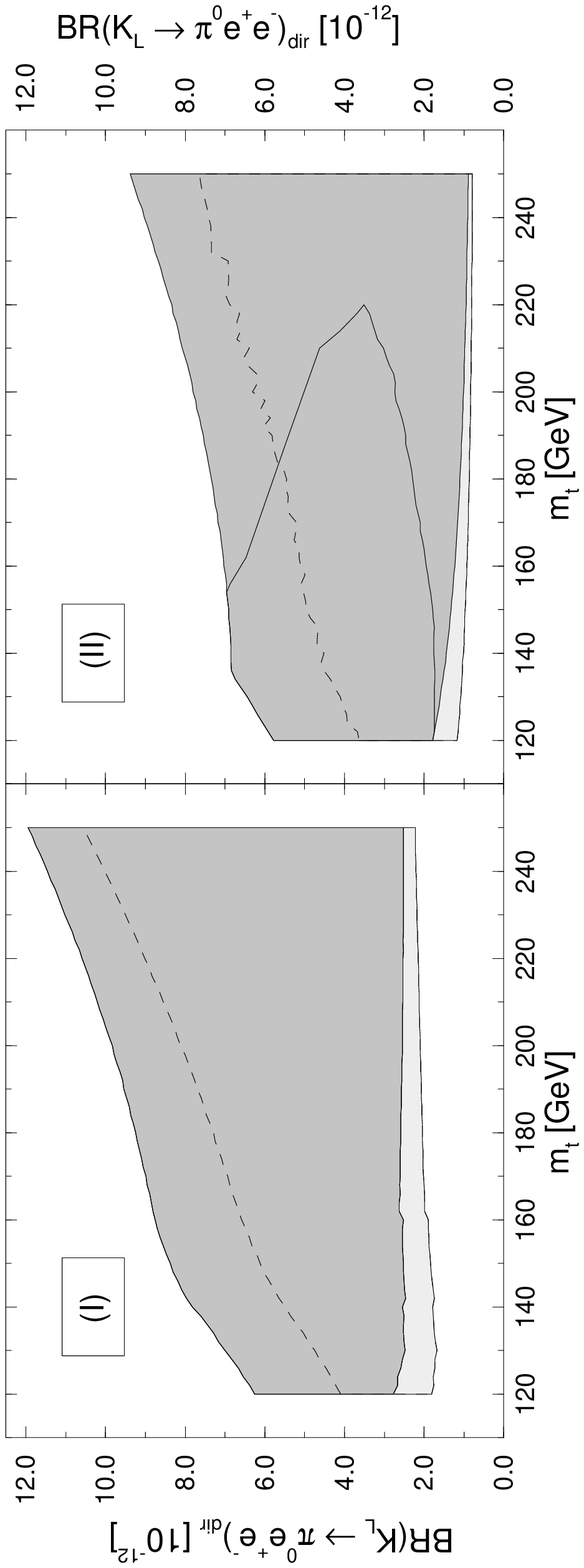}
} }
\vspace{0.15in}
}{
\caption[]{
Allowed ranges for $BR(K_L \longrightarrow \pi^0 e^+e^-)_{\rm dir}$ as
a function of $m_t$ for ${\Lambda_{\overline{\rm MS}}}=0.3\gev$ at
scale $\mu=1.0\gev$. The labels (I) and (II) refer to the two possible
solutions for the CKM phase $\delta$. Parameters $|V_{cb}|$,
$|V_{ub}/V_{cb}|$ and $B_K$ were varied within present experimental
limits as given in (\ref{rangea}).  The light and dark shaded areas
correspond to the LO and NLO results, respectively. The solid line
inside the dark shaded area describes the additional restriction from
$B^0$--$\overline{B}^0$ mixing (see the text).
\label{fig:1}}
\end{figure}
}

\figatend{\begin{figure}[h]}{
\vspace{0.15in}
\centerline{
\epsfysize=6.5in
\rotate[r]{
\epsffile{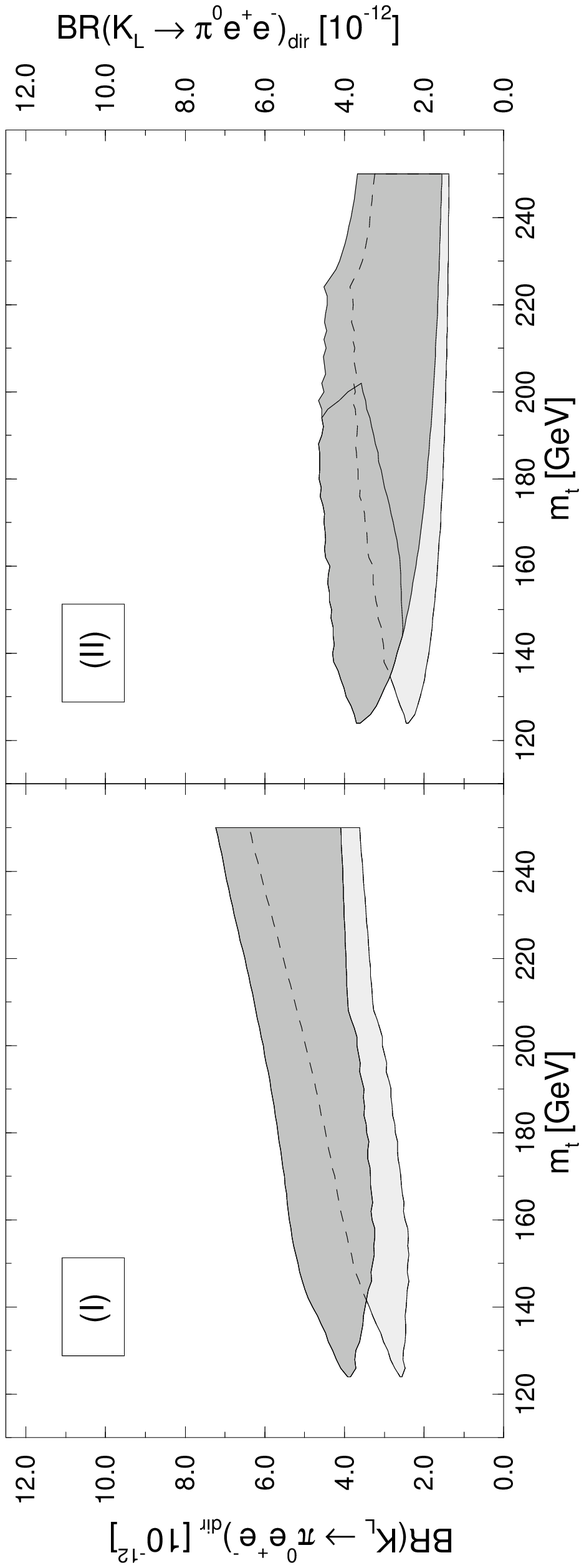}
} }
\vspace{0.15in}
}{
\caption[]{
The same as in fig.~\ref{fig:1}, but for the parameter range given in
(\ref{rangeb}).
\label{fig:2}}
\end{figure}
}

\figatend{\begin{figure}[h]}{
\vspace{0.15in}
\centerline{
\epsfysize=6.5in
\rotate[r]{
\epsffile{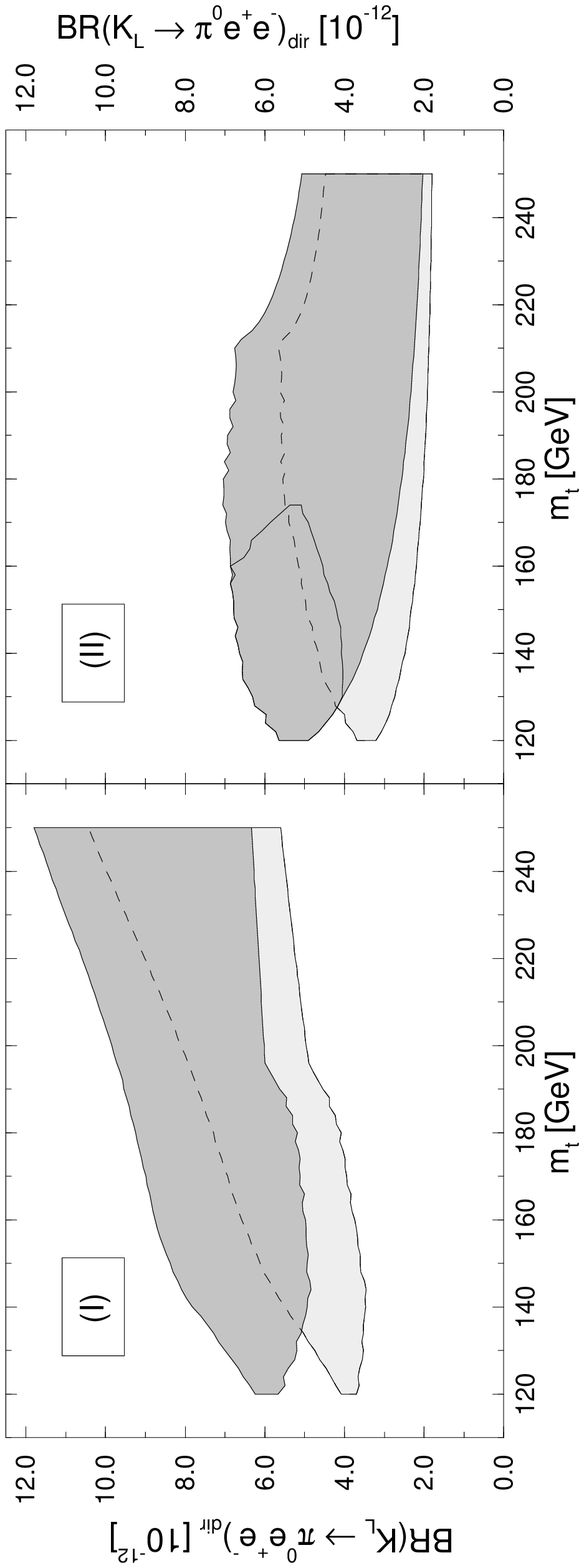}
} }
\vspace{0.15in}
}{
\caption[]{
The same as in fig.~\ref{fig:1}, but for the parameter range given in
(\ref{rangec}).
\label{fig:3}}
\end{figure}
}

\figatend{\begin{figure}[h]}{
\vspace{0.15in}
\centerline{
\epsfysize=4in
\rotate[r]{
\epsffile{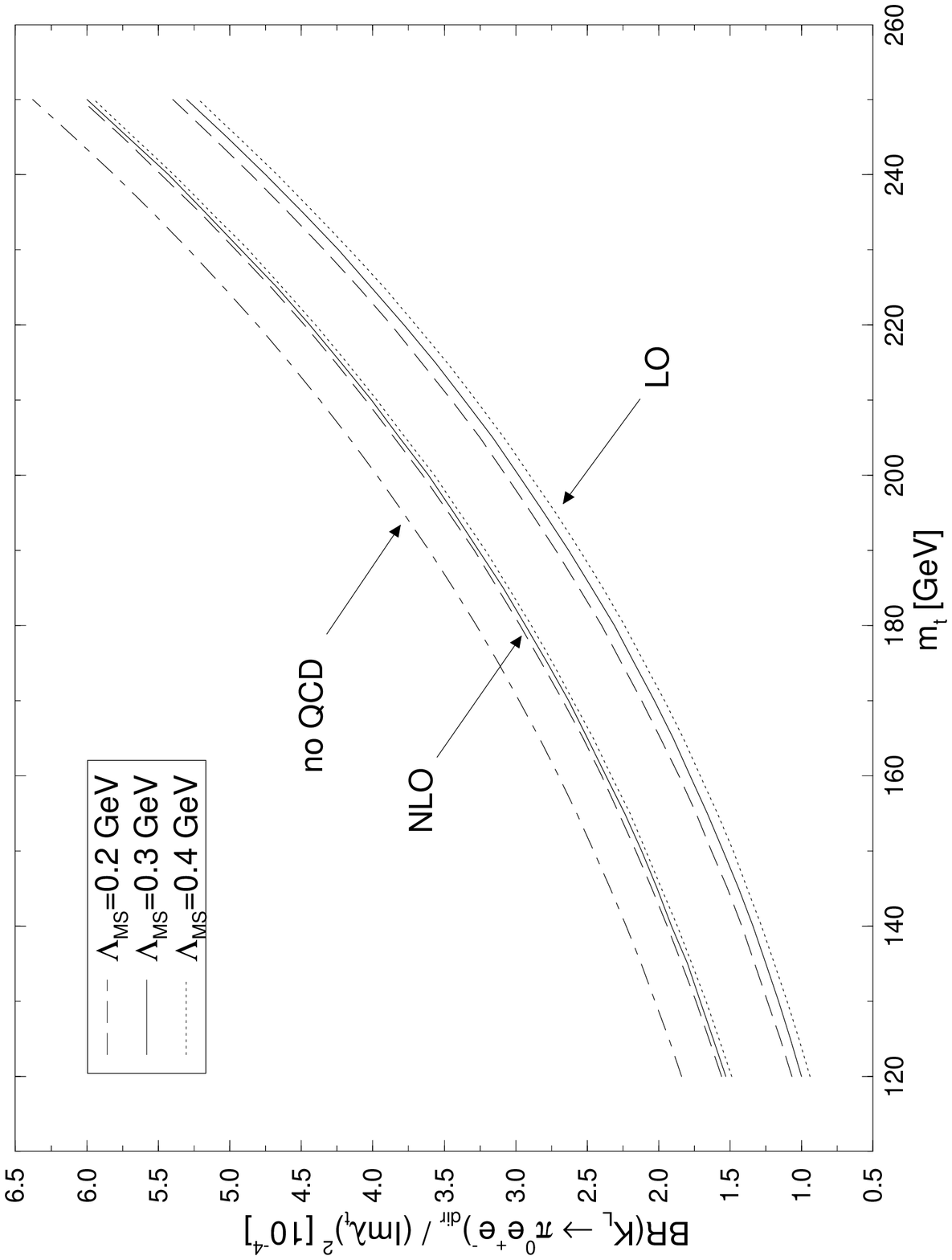}
} }
\vspace{0.15in}
}{
\caption[]{
$BR(K_L \longrightarrow \pi^0 e^+e^-)_{\rm dir}/(\IM\lambda_t)^2$
as a function of $m_t$ for various values of ${\Lambda_{\overline{\rm
MS}}}$ at scale $\mu=1.0\gev$.
\label{fig:4}}
\end{figure}
}

\end{appendix}

\end{document}